\documentclass[a4paper,11pt]{article}
\pdfoutput=1 

\usepackage{jcappub} 

\usepackage[T1]{fontenc} 
\usepackage{graphicx}
\usepackage{epsfig}
\usepackage{rotate}
\usepackage{amsmath}
\usepackage{amssymb}
\usepackage{amsfonts}
\usepackage{multirow}
\usepackage{tablefootnote}
\usepackage{diagbox}
\usepackage{slashbox}
\usepackage{bm}
\usepackage{enumerate}
\usepackage{afterpage}
\usepackage{xcolor}
\usepackage{natbib, hyperref}
\usepackage{cancel}

\def\be{\begin{equation}}
\def\ee{\end{equation}}

\def\bea{\begin{eqnarray}}
\def\eea{\end{eqnarray}}

\def\k{\bm{k}}
\def\v{\bm{v}}
\def\r{\bm{r}}
\def\rh{\hat{\bm{r}}}
\def\H{{\cal H}}
\def\d{\mathrm{d}}

\newcommand{\ud}{\mathrm{d}}
\newcommand{\p}{\partial}
\newcommand{\cH}{\mathcal{H}}
\newcommand{\cl}{\mathcal{L}}
\newcommand{\Q}{\mathcal{Q}}
\newcommand{\I}{\mathrm{i}}
\newcommand{\e}{\mathrm{e}}

\newcommand{\redd}[1]{{\color{black}{#1}}}
\newcommand{\red}[1]{{\color{black}{#1}}}

\newcommand{\blue}[1]{{\color{blue}{#1}}}

\newcommand{\CC}[1]{{\color{black}{#1}}}

\allowdisplaybreaks

\def\Xint#1{\mathchoice {\XXint\displaystyle\textstyle{#1}}% 
{\XXint\textstyle\scriptstyle{#1}}%
 {\XXint\scriptstyle\scriptscriptstyle{#1}}%
 {\XXint\scriptscriptstyle\scriptscriptstyle{#1}}%
  \!\int} \def\XXint#1#2#3{{\setbox0=\hbox{$#1{#2#3}{\int}$} \vcenter{\hbox{$#2#3$}}\kern-.5\wd0}} \def\ddashint{\Xint=} \def\dashint{\Xint-}

\title{{Wide-angle effects in multi-tracer power spectra with Doppler corrections}}

\author{Pritha Paul$^{1}$, Chris Clarkson$^{1,2}$, and Roy Maartens$^{2,3,4}$}
\affiliation{ $^{1}$School of Physics \& Astronomy, Queen Mary University of London, London E1 4NS, UK \\
$^2$Department of Physics \& Astronomy, University of the Western Cape, Cape Town 7535, South Africa\\
$^{3}$Institute of Cosmology \& Gravitation, University of Portsmouth, Portsmouth PO1 3FX, UK\\
$^4$National Institute for Theoretical and Computational Sciences (NITheCS), Cape Town 7535, South Africa
}

\abstract{We examine the computation of wide-angle corrections to the galaxy power spectrum including redshift-space distortions and relativistic Doppler corrections, and also including multiple tracers with differing clustering, magnification and evolution biases. We show that the inclusion of the relativistic Doppler contribution\CC{, as well as radial derivative terms, are} crucial for a consistent wide-angle expansion for large-scale surveys, both in the single and multi-tracer cases.  We also give for the first time the wide-angle cross-power spectrum associated with the Doppler magnification-galaxy cross correlation, which has been shown to be a new way to test general relativity.  
In the full-sky power spectrum, the wide-angle expansion allows integrals over products of spherical Bessel functions to be computed analytically as distributional functions, which are then relatively simple to integrate over. 
We give for the first time a complete discussion and new derivation of the finite part of the divergent integrals of the form $\int_{0}^{\infty} \mathrm{d} r r^{n} j_{\ell}(k r) j_{\ell^{\prime}}(q r)$, which are necessary to compute the wide-angle corrections when a general window function is included. This facilitates a novel method for integrating a general analytic function against a pair of spherical Bessel functions. 
}

\begin{document}
\maketitle
\date{\today}
\flushbottom

\section{Introduction}

The two-point correlation function (2PCF) in observed (redshift) space is often expressed in the plane-parallel or flat-sky approximation, in which the directions from the observer to galaxy pairs are assumed to be nearly equal, $\rh_1\approx \rh_2$. Galaxy surveys with wide sky coverage, in particular next-generation surveys, require us to move beyond the flat-sky limit and include wide-angle correlations, with $\rh_1\neq \rh_2$. This was shown in early work by  \cite{Szalay:1997cc,Matsubara:1999du} (using a  tripolar spherical harmonic expansion) and then further investigated in, e.g., \cite{Szapudi:2004gh, Papai:2008bd, Raccanelli:2010hk,
Bertacca:2012tp,Yoo:2013zga, Reimberg:2015jma, Raccanelli:2016avd,Tansella:2017rpi,Castorina:2017inr,Beutler:2018vpe,Beutler:2021eqq,Castorina:2021xzs,Noorikuhani:2022bwc}. Our aim is to expand and clarify a number of these results to the multi-tracer power spectrum, including relativistic corrections. \CC{We give for the first time the power spectrum and wide angle corrections to the cross-correlation between the galaxy number counts and the Doppler magnification induced by peculiar velocities.  We also include contributions from the derivatives of the growth rate and clustering biases, which are neglected in other studies yet, as we show, are needed for a consistent analysis on large scales.}

\subsection{\redd{Doppler contribution to the number counts}}

{The galaxy number density contrast at the source is $\delta_g=(n_g-\bar{n}_g)/\bar{n}_g=b\,\delta_{\rm m}$, where $b$ is the clustering bias and $\delta_{\rm m}$ is the matter density contrast. For brevity, we write $\delta_g \equiv \delta$ and omit the $z$-dependence in our expressions. At first (linear) order in perturbations, the number density contrast at the source is related to the  contrast $\delta^s$ that is observed in redshift space by}
\be \label{1a}
{\delta^s(\r_i) = \delta(\r_i) -{1\over \H_i}{\p  \over \p r_i}\big(\v_i\cdot \rh_i\big)-{\alpha_i\over \H_i}\,\big(\v_i\cdot \rh_i\big)\,,}
\ee
where we use Newtonian gauge.
Here $\v_i=\v(\r_i)$ is the peculiar velocity, $\r_i=r(z_i)\rh_i$, where $r$ is the comoving line-of-sight distance, and $\H_i=\H(z_i)$ is the conformal Hubble rate. The second term on the right of \eqref{1a} is the standard Kaiser redshift-space distortion (RSD), while the third term is a Doppler redshift effect. This Doppler term is  suppressed relative to the Kaiser RSD term by a factor $\cH/k$ in Fourier space. It is therefore typically neglected in most work on galaxy clustering in redshift space.

The Doppler coefficient in \eqref{1a} is given on a spatially flat background by  
(e.g. \cite{Challinor:2011bk,Bertacca:2012tp};  see \cite{DiDio:2016ykq} for the generalisation to spatially curved backgrounds):
\be  \label{4}
\alpha_i = {2\over r_i}  - \H_i\, b_{{\rm e}\,i} -{\ud{\H_i} \over \ud\ln(1+z_i)} +2 \H_i\,\Q_i \left(1-{1 \over r_i\H_i} \right).
\ee
In the original RSD paper \cite{Kaiser:1987qv},  the Doppler coefficient \eqref{4} includes only the first 2 terms on the right. This is 
followed in the pioneer wide-angle papers \cite{Szalay:1997cc,Matsubara:1999du} and many subsequent papers. However, the second 2 terms on the right of \eqref{4} are required for a correct analysis \cite{Challinor:2011bk,Bertacca:2012tp}.
In \eqref{4},  the `evolution bias' $b_{\rm e}= -\p\ln \bar{n}_g/\p\ln(1+z)$ measures the deviation of the average comoving number density from constancy (due e.g. to galaxy mergers) \cite{Maartens:2021dqy}.
The third term on the right takes account of cosmic evolution.
The last term on the right arises from the Doppler correction to lensing convergence \cite{Bonvin:2005ps,Bonvin:2008ni} in a flux-limited survey,
where $\Q=-\p\ln \bar{n}_g/\p\ln L_{\rm c}$ is the magnification bias and $L_{\rm c}$ is the luminosity cut \cite{Maartens:2021dqy}. (In the ideal case of no flux limit, $\Q=0$, and for line intensity mapping, $\Q=1$.)

In \eqref{1a} we have omitted two contributions to the number density contrast:
\begin{enumerate} 
\item[(1)]
the  contribution from the standard lensing magnification term $2(\Q-1)\kappa$, where $\kappa$ is a weighted integral of $\delta_{\rm m}$ along the line of sight; 
\item[(2)]
additional relativistic potential terms, including Sachs-Wolfe, integrated Sachs-Wolfe and time delay effects, which collectively scale as  the gravitational %(Bardeen) 
potential $\Phi$. 
\end{enumerate}

The standard lensing magnification contribution (1) is typically only important at higher redshift -- and it requires significant additional complexity to incorporate it into the Fourier power spectrum.
The additional relativistic terms in (2) are suppressed relative to the Doppler term in \eqref{1a}, since the Poisson equation shows that $\Phi\sim (\H^2/k^2)\delta_{\rm m}$.  (See e.g. \cite{Challinor:2011bk,Bertacca:2012tp,Tansella:2017rpi} for details of all these terms.)

There is a subtle point about the Doppler contribution relative to the potential contribution. In $\delta^s$, the Doppler term is clearly less suppressed than the potential contribution.  However, this does not translate directly to the 2PCF in the case of a single tracer with correlations at equal redshifts. In this case, i.e. auto-correlations at equal $z$, the Doppler contribution  is a square of the Doppler term, i.e., scaling as $(\cH^2/k^2)P_{\rm m}$, like the leading potential contribution. \red{Strictly, this means that it is inconsistent to neglect the  potential contributions while including the Doppler term, when considering auto-correlations at equal redshifts. Including the potential terms is simple in principle, but we omit them in order to avoid additional complexity in the equations. Effectively, this means that we adopt a `weak field' approximation \cite{DiDio:2018zmk}. }

When considering  correlations of two tracers (see e.g. \cite{McDonald:2009ud, Bonvin:2013ogt, Bacon:2014uja, Gaztanaga:2015jrs,Hall:2016bmm,Lepori:2017twd,Breton:2018wzk, DiDio:2018zmk,DiDio:2020jvo,Beutler:2020evf,Beutler:2021eqq}), the leading Doppler contribution to the 2PCF scales instead as $(\H/k)P_{\rm m}$ -- and in this case it is consistent to neglect the potential contributions. Note that it is also consistent in the case of single-tracer correlations at unequal redshifts.

\begin{figure}[h]
\centering
\includegraphics[width=.4\textwidth]{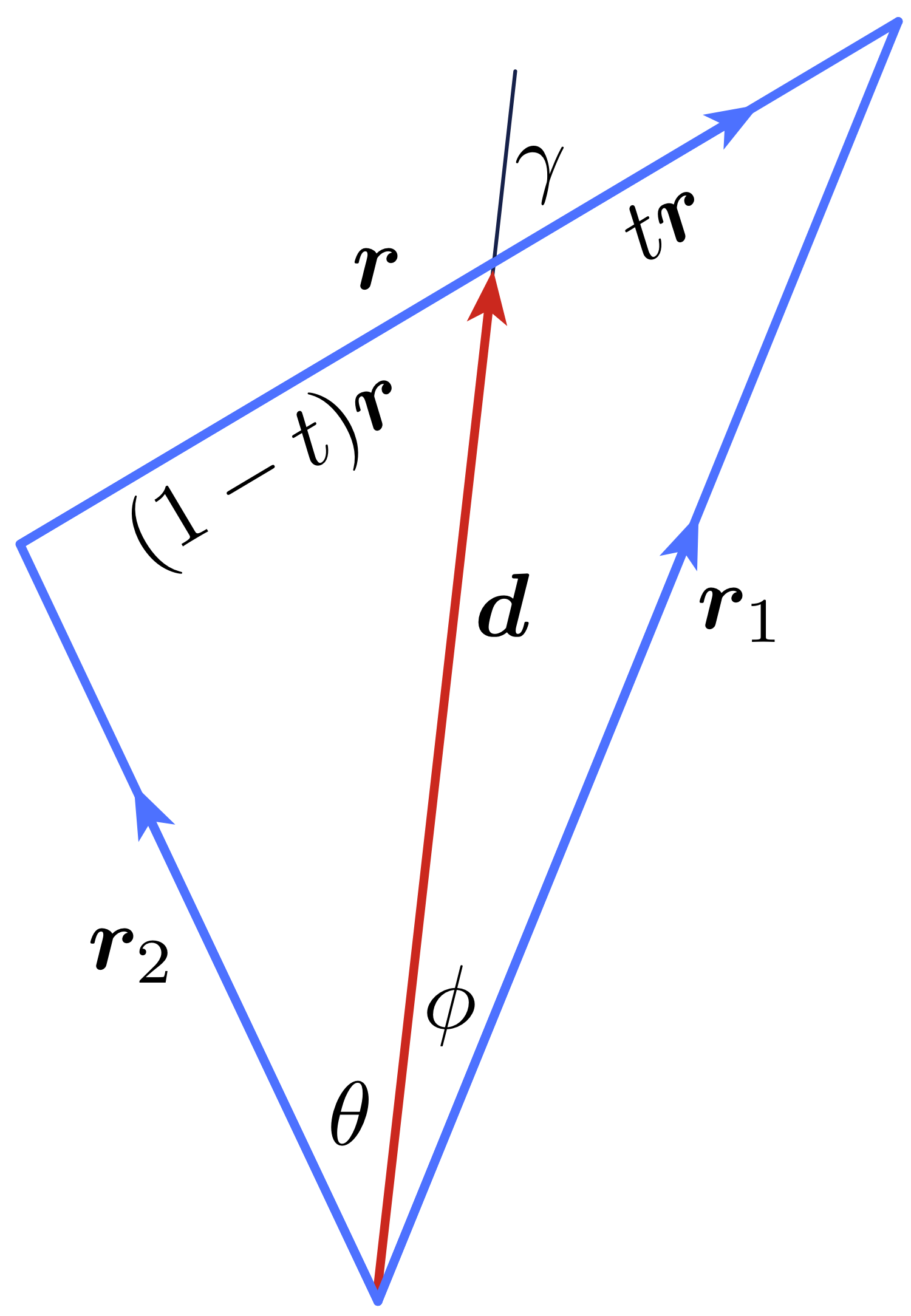}
%\includegraphics[width=.49\textwidth]{geometry.png}
%\vspace*{-0.5cm}
\caption{Geometry of the galaxy pair and observer.
%Note that $0\leq \theta\leq \pi/2$. Also, for $\theta>0$, we have $\theta\leq \gamma\leq \pi-\theta$.
}  \label{fig1}
\end{figure}

{Furthermore, we highlight the fact that the leading wide-angle contribution to $\delta^s$ scales as $r/d \sim d^{-1}/k$, where $r$ is the comoving separation of the galaxy pair and $d$ is the line-of-sight comoving distance to the galaxy pair (see \autoref{fig1} and \autoref{s1.1}). As a consequence, the leading wide-angle and leading Doppler contributions are of the same order  -- so that a consistent treatment requires the inclusion of both (e.g. \cite{Castorina:2021xzs,Noorikuhani:2022bwc}).} 
\CC{In addition to this, for a consistent treatment we need to include radial derivatives of the growth rate, the biases, and other variables which appear. In an expansion in $r/d$, derivative terms appear at order $\sim r{\cal H}$ when approaching $z\sim 1$~-- i.e.,  distances of the Hubble scale to the galaxy pair -- and so are also needed for a consistent treatment.  In both cases these have been neglected in previous analyses.}

We define the transforms \cite{Szalay:1997cc,Matsubara:1999du,Bertacca:2012tp}
\be
A^n_\ell(\r) = \int {\ud^3\k \over (2\pi)^3}\,(\I k)^{-n}\cl_\ell(\hat\k\cdot\hat\r) \,\e^{\I \k\cdot\r}\,\delta_{\rm m}(\k),
\label{7}
\ee
where $\cl_\ell$ is a Legendre polynomial,
and then we can express \eqref{1a} as
\be
{\delta^s(\r_i)\over b_i} = \Big(1+ { \beta_i\over 3} \Big)A^0_0(\r_i)+ {2\over3}\beta_i\, A^0_2(\r_i)  +\beta_i\,\alpha_i \,A^1_1(\r_i)  \quad
\mbox{where}\quad\beta_i\equiv {f_i\over b_i}.
 \label{6}
\ee
It follows that the 2PCF in redshift space, $\xi_{{g}}(\r_1,\r_2)=\langle \delta^s(\r_1)\,\delta^s(\r_2)\rangle $, is given by
\begin{align}
{\xi_{{g}}(\r_1,\r_2)\over b_1b_2} =& \Big(1+ { \beta_1\over 3} \Big)\Big(1+ { \beta_2\over 3} \Big)S^0_{00}(\r_1,\r_2)+ {4\over9}\beta_1\beta_2\, S^0_{22}(\r_1,\r_2)  +
{2\beta_2\over3}\Big(1+ { \beta_1\over 3} \Big)S^0_{02}(\r_1,\r_2) 
\notag\\
& +{2\beta_1\over3}\Big(1+ { \beta_2\over 3} \Big)S^0_{20}(\r_1,\r_2)
+\Big(1+ { \beta_1\over 3} \Big)\beta_2\alpha_2 S^1_{01}(\r_1,\r_2)+\Big(1+ { \beta_2\over 3} \Big)\beta_1\alpha_1  S^1_{10}(\r_1,\r_2)
\notag\\
&
+{2\over3}\beta_1\beta_2\Big[\alpha_1S^1_{12}(\r_1,\r_2)+ \alpha_2S^1_{21}(\r_1,\r_2)\Big]
+\beta_1\beta_2\,\alpha_1\alpha_2 S^2_{11}(\r_1,\r_2), \label{8}
%\label{djksndksnjcdskjdn}
\end{align}
where
\bea
S^{n_1+n_2}_{\ell_1\ell_2}(\r_1,\r_2) \equiv (-1)^{\ell_2} \int {\ud^3\k \over (2\pi)^3}\,(\I k)^{-(n_1+n_2)}\,\cl_{\ell_1}(\hat\k\cdot\hat\r_1)\,\cl_{\ell_2}(\hat\k\cdot\hat\r_2) \,\e^{\I \k\cdot(\r_1-\r_2)}\,P_{\rm m}(k).
\label{9}
\eea

Thus in general the 2PCF is a function of $\bm r_1$ and $\bm r_2$, or equivalently of
\be
\bm r=\r_1-\bm r_2 \quad \mbox{and} \quad \bm d = (1-t)\r_1+t\r_2\quad (0\leq t\leq 1),
\ee 
where $t$ determines the choice of $\bm d$ --  see \autoref{fig1}. 
%We define the wide-angle redshift-space power spectrum at a displacement $\bm{d}$ from the observer by Fourier transforming the redshift-space 2PCF over the vector connecting the two galaxies:
%\be
%P_{{g}}(\bm d, \bm k)=\int \ud^3\bm r \,{\rm e}^{-\I\,\k\cdot\r}\,\xi_{{g}}(\bm d,\bm r) \,.
%\ee

We can also define the wide-angle 2-point cross-correlation of galaxy number density contrast and Doppler magnification $\xi_\kappa(\bm r_1,\bm r_2)=\big\langle \delta^s(\r_1)\,\kappa_v(\r_2)\big\rangle$, where the magnification induced by peculiar velocities is given by \cite{Bacon:2014uja}:
\bea
\kappa_v =-{\tilde \alpha \over \cH}\, \bm v\cdot\bm n\qquad\mbox{where}\quad \tilde \alpha={1\over r}-{\cH}\,.
\eea
Then the equivalent of \eqref{8} becomes
\begin{align}
{\xi_\kappa(\bm r_1,\bm r_2)\over b_1f_2\tilde\alpha_2} =& \Big(1+ { \beta_1\over 3} \Big) S^1_{01}(\r_1,\r_2)
+{2\over3}\beta_1\,S^1_{21}(\r_1,\r_2)
+\beta_1\alpha_1\, S^2_{11}(\r_1,\r_2)\,. \label{8dop}
\end{align}
As shown in \cite{Bonvin:2016dze}, this 2-point correlation function has a significant dipole which can be used as a test of general relativity~\cite{Andrianomena:2018aad,Franco:2019wbj}. The power spectrum of this 2PCF has not been given before. 

\subsection{Wide-angle multipole expansion -- overview}
\label{s1.1}

Here we give a brief summary of the calculation of the wide-angle expansion. 
In general the multipole decomposition of $\xi_{{g}}(\bm d,\bm r)$ and $\xi_\kappa(\bm d,\bm r)$  or their equivalent power spectra $P_{{g}}(\bm d, \bm k)$ and $P_\kappa(\bm d, \bm k)$ (defined below) are analytically intractable. We aim to produce a series expansion first in the 2PCF about the plane-parallel limit ($r\ll d$), and then for each term in that series expansion, we perform a multipole decomposition with respect to the angle between $\r$ and $\bm d$, with cosine $\mu=\hat{\bm d}\cdot{\hat \r}$. Once translated to the power spectrum this becomes an expansion about $k^{-1}\ll d$ with coefficients expanded in Legendre multipoles in $\mu_k=\hat{\bm d}\cdot{\hat \k}$. Clearly a sensible expansion variable in the 2PCF is $x=r/d$, but we need to be careful. The plane-parallel limit is not simply the limit as $r\to0$ with $d$ fixed:  in this limit, $\xi_{\rm pp}(r)\sim \int \d k j(kr)P_m(k)$, and therefore $\xi_{\rm pp}$ is a function of $r$. Nor is it the limit $d\to\infty$ with $r$ fixed -- since the coefficients $\alpha$ and $\beta$ are functions of the two redshift shells $\bm r_1,\bm r_2$ we are looking at [see \eqref{skjncskjdbvf} below]. 
The plane-parallel limit is thus  a mixture of both $r\to0$ and $d\to\infty$. 

On inspection each term in \eqref{8} contains parts which depend on $|\r_1|,|\r_2|$~-- i.e., the $\alpha$ and $\beta$ coefficients only depend on the distance to each source,  and parts which depend on $\hat\r_1,\hat\r_2,\hat\r$ and $|\r|$ but not on $|\r_1|,|\r_2|$~-- i.e., the terms which are integrals over the power spectrum, $S^n_{\ell_1,\ell_2}$, depend on the geometry of the triangle and the distance between the sources. These different contributions require slightly different series expansions around the plane-parallel limit:
\begin{itemize}
    \item 
In $S^n_{\ell_1,\ell_2}$, we may expand in a series in $x=r/d$ around $x=0$ with $r$ fixed, because the direction vectors $\hat\r_1,\hat\r_2,\hat\r$ only depend on the ratio $x$, not on $r$ and $d$ separately. In addition,  $|\r|$ in the exponential does not depend on $\mu$:  hence it does not affect the multipoles and does not need expanding (the $\mu$ dependence factors into Legendre polynomials on using a plane wave expansion below). 
\item
Functions of $|\r_1|,|\r_2|$ which appear can sensibly be expanded in a series in $x$, but now with $d$ fixed, so that the series coefficients come out in terms of  $\alpha(d)$ and $\beta(d)$ and their derivatives  $\d\alpha(d)/\d\ln d$, $\d\beta(d)/\d\ln d$ evaluated at $r=0$. We start with a general $\bm r_1\neq \bm r_2$ and then expand functions of these quantities around a median distance given by the shell $d$ [see, e.g., \eqref{cjdhsbdjsbdjsbc}]. 
\end{itemize}

Putting this all together gives a multipole series of the form,
\be\label{dksjndckdncs}
\xi_{{g}}(\bm d,\bm r) = \sum_{\ell,p} \,\Xi_\ell^{(p)}(r,d) \left(\frac{r}{d}\right)^p \mathcal{L}_\ell(\mu)\quad \mbox{where}\quad \mu=\hat{\bm d}\cdot{\hat \r}=\cos\gamma\,,
\ee
and the $\ell$-pole is given by
\be 
\sum_{p} \,\Xi_\ell^{(p)}(r,d) \left(\frac{r}{d}\right)^p\,. \label{e12}
\ee 
We will denote analogous coefficients of the galaxy-magnification 2PCF with a tilde. 
The plane-parallel approximation $p=0$ leads to the well-known coefficients for the galaxy-galaxy power spectrum for two tracers. The leading terms at order $(\cH/k)^0$ are the set of even multipoles given first in~\cite{Szalay:1997cc,Matsubara:1999du},
\begin{align}
\Xi^{(0)}_{0}&=\frac{1}{15}b_1b_2 \Big(15+5{\beta_{1}}+5{\beta_{2}}+3{\beta_{1} \beta_{2}}\Big) \xi^{(0)}_{0}, \\
%\Xi^{(0)}_{1}&=\frac{b_1b_2}{5}\Big[5\beta_{1} \alpha_{1}-5\beta_{2}\alpha_{2} +{3 \beta_{1} \beta_{2}\big( \alpha_{1}}-{ \alpha_{2}}\big) \Big] \xi^{(1)}_{1}, \\
\Xi^{(0)}_{2}&=-\frac{2}{21}b_1b_2\Big({7 \beta_{1}}+{7 \beta_{2}}+{6 \beta_{1} \beta_{2}}\Big) \xi^{(0)}_{2}, \\
%\Xi^{(0)}_{3}&=\frac{2b_1b_2}{5}\, \beta_{1} \beta_{2} \big(\alpha_{2}-\alpha_{1}\big) \xi^{(1)}_{3},\\
\Xi^{(0)}_{4}&=\frac{8}{35}b_1b_2\, \beta_{1} \beta_{2} \,\xi^{(0)}_{4}\,.
\end{align}
At order $(\cH/k)^1$  we have the sub-leading odd multipoles arising from the Doppler contribution given here for the first time,
\begin{align}
%\Xi^{(0)}_{0}&=\frac{b_1b_2}{15} \Big(15+5{\beta_{1}}+5{\beta_{2}}+3{\beta_{1} \beta_{2}}\Big) \xi^{(0)}_{0}, \\
\Xi^{(0)}_{1}&=\frac{1}{5}b_1b_2\Big[5\beta_{1} \alpha_{1}-5\beta_{2}\alpha_{2} +{3 \beta_{1} \beta_{2}\big( \alpha_{1}}-{ \alpha_{2}}\big) \Big] \xi^{(1)}_{1}, \\
%\Xi^{(0)}_{2}&=-\frac{2b_1b_2}{21}\Big({7 \beta_{1}}+{7 \beta_{2}}+{6 \beta_{1} \beta_{2}}\Big) \xi^{(0)}_{2}, \\
\Xi^{(0)}_{3}&=\frac{2}{5}b_1b_2\, \beta_{1} \beta_{2} \big(\alpha_{2}-\alpha_{1}\big) \xi^{(1)}_{3}\,.
%\Xi^{(0)}_{4}&=\frac{8b_1b_2}{35}\, \beta_{1} \beta_{2} \,\xi^{(0)}_{4}\,.
\end{align}
Note these vanish in the case of a single tracer.
The coefficients $\Xi_\ell^{(p)}$ of the series in \eqref{e12} depend on $d$ via the coefficients $\alpha,\beta$ and on the separation $r$ via a sum over weighted integrals of the power spectrum:
\bea
\xi^{(n)}_{\ell'}(r) &\equiv & \int {\ud k \over 2\pi^2}\, k^{2-n}j_{\ell'}(kr) P_{\rm m}(k).
\label{12}
\eea
Note that these terms are of order $(1/k)^n$. 

In the case of the galaxy-magnification power spectrum the plane-parallel limit has a dipole and an octupole at order $n=1$, corresponding to terms $\sim(\cH/k)^1$ :
\begin{align}
\tilde\Xi^{(0)}_{1}&=-\frac{1}{5}b_{1}f\tilde\alpha_{2} { \big(5+3 \beta_{1}\big)\xi^{(1)}_{1}}\,, \\
%\Xi_{0,2}=0 \\
\tilde\Xi^{(0)}_{3}&=\frac{2}{5}{b_{1} f \beta_{1}  \tilde\alpha_{2} \, \xi^{(1)}_{3}} \,.
%\Xi_{0,4}=0
\end{align}
At order $(\cH/k)^2$ ($n=2$), we have the sub-leading corrections to the monopole and quadrupole:
\begin{align}
\begin{gathered}
\tilde\Xi^{(0)}_{0}=-\frac{1}{3}{ b_{1} f \beta_{1} \alpha_{1}\tilde\alpha_{2} \xi^{(2)}_{0}} \\
\tilde\Xi^{(0)}_{2}=\frac{2}{3}{ b_{1} f \beta_{1} \alpha_{1}\tilde\alpha_{2}  \xi^{(2)}_{2}}
\end{gathered}
\end{align}
We will derive the other coefficients below. 
Only in the plane-parallel approximation is the $\ell$-pole dependent solely on $\xi^{(n)}_{\ell}(r)$~-- in general, differing $\ell'$-poles, $\xi^{(n)}_{\ell'\neq\ell}(r)$, come into play, and \eqref{e12} leads to:
\be\label{djschsbdisuhdishd}
\Xi_\ell^{(p)}(r,d) = \sum_{\ell',n} \Xi_{\ell\ell'}^{(p,n)}(d)\xi^{(n)}_{\ell'}(r)\,.
\ee
Here $\Xi_{\ell\ell'}^{(p,n)}(d)$ are functions of $\alpha(d)$, $\beta(d)$ and their derivatives. 

Once we have the 2PCF in the form~\eqref{dksjndckdncs}, we define the wide-angle  power spectrum at a displacement $\bm{d}$ from the observer by Fourier transforming the redshift-space 2PCF over $\bm{r}$ (see \autoref{fig1}):\footnote{Note that this is treated as a formal Fourier transform over $r\in[0,\infty)$, not as a discrete Fourier series over a finite $r$. A window function can be added to account for such effects. }
\be\label{psdk}
P_{{g}}(\bm d, \bm k)=\int \ud^3 \r \,{\rm e}^{-\I\,\k\cdot\r}\,\xi_{{g}}(\bm d,\bm r)\,.
\ee
The power spectrum can be expanded in multipoles defined by the angle between the wavevector $\k$ and the line of sight $\bm d$, as 
\be
P_{{g}}(\bm d, \bm k) = \sum_\ell \mathcal{P}_\ell(k,d)\, \cl_\ell(\mu_k)
\quad\mbox{where}\quad {\mu_k=\hat{\bm d}\cdot\hat{\bm k}}\,.
\ee
Then, using the plane-wave expansion
\be
{\rm e}^{-\I\,\k\cdot\r} = \sum_{\ell=0}^\infty \I^{-\ell} (2\ell+1) j_\ell(kr)
\cl_\ell(\hat\k\cdot\hat\r)\,,
\ee
we find that
\be
\mathcal{P}_\ell(k,d) = 4\pi\, \I^{-\ell} \sum_p\int \d r\,r^2\,
\Xi_\ell^{(p)}(r,d)\, j_\ell(kr)\left(\frac{r}{d}\right)^p \,.
\ee

An important difference in the multipoles in redshift space versus Fourier space is that in redshift space the multipoles are in $\mu$ with $r$ fixed, while in Fourier space the multipoles are in $\mu_k$ with $k$ fixed. 
On using~\eqref{12} and \eqref{djschsbdisuhdishd}, we  find that the multipoles become 
\be\label{csdcndkncsk}
\mathcal{P}_\ell(k,d) = \frac{2}{\pi}\,\I^{-\ell}\sum_{\ell', n, p} d^{-p}\,\Xi_{\ell\ell'}^{(p,n)}(d)\int {\ud q}\, q^{2-n}\,P_{\rm m}(q)\, \mathcal{I}^p_{\ell\ell'}(k,q)\,,
\ee
where 
\be\label{ipll}
\mathcal{I}^p_{\ell\ell'}(k,q)=\int_0^\infty \ud r \,r^{2+p}\, j_\ell(kr)\,j_{\ell'}(qr)\,.
\ee
These integrals, although divergent, can be evaluated as distributions~-- delta functions, and more complicated singular points~-- which then simply feed into the integral over the power spectrum in \eqref{csdcndkncsk}.
We can rewrite \eqref{csdcndkncsk} as a series in $k^{-1}/d=(kd)^{-1}$ (which is a Fourier counterpart to $r/d$):
\be
\mathcal{P}_\ell(k,d)=\sum_{p} {\mathcal{P}^{(p)}_\ell {(k,d)}}\,{(kd)^{-p}}\,,
\ee
with
\begin{align}
\mathcal{P}^{(p)}_\ell(k,d)
&=\frac{2}{\pi}\I^{-\ell}\sum_{\ell', n}{\Xi_{\ell\ell'}^{(p,n)}(d)}\,P^{pn}_{\ell\ell'}(k)\,,\label{cppl}
\end{align}
where 
%[NOTE CHANGE OF DEFINITION TO INCLUDE $k$]
\be \label{pnpll}
P^{pn}_{\ell\ell'}(k)=k^p\int_0^\infty {\ud q}\, q^{2-n}\,P_{\rm m}(q)\, \mathcal{I}^p_{\ell\ell'}(k,q)\,.
\ee

\section{Evaluating the 2PCF and power spectrum}

In order to evaluate the 2PCF, we first evaluate $S^{n_1+n_2}_{\ell_1\ell_2}(\r_1,\r_2)$ in \eqref{9} finding, %using~\cite{Szalay:1997cc}
%\begin{align}
%S_{\ell_{1} \ell_{2}}^{n}=(4 \pi)^{3 / 2}(-1)^{\ell_{1}} \sum_{L} \I^{L-n} B_{\ell_{1} \ell_{2}}^{L}(\Delta)\, \xi_{L}^{(n)}(r)\,,
%\end{align}
\begin{align}
    S_{\ell_{1} \ell_{2}}^{n}=&\sum_{L}(-1)^{\ell_2} i^{L-n} (2L+1)\left[\frac{(4\pi)^3(2L + 1)}{(2\ell_1 + 1)(2\ell_2 +1 )}\right]^{1/2}
    \left(\begin{array}{lll}
\ell_{1} & \ell_{2} & L \\
0 & 0 & 0
\end{array}\right) 
\xi_{L}^{(n)}(r)
\notag \\
&\sum_{m_1,m_2,M}
 \left(\begin{array}{lll}
\ell_{1} & \ell_{2} & L \\
m_1 & m_2 & M
\end{array}\right) 
Y_{\ell_{1} m_{1}}\left(\hat{\boldsymbol{r}}_{1}\right) Y_{\ell_{2} m_{2}}\left(\hat{\boldsymbol{r}}_{2}\right)Y_{L M}(\hat{\boldsymbol{r}})
\end{align}
\iffalse
where the triangle-shape coefficients are
\begin{align}\label{dsjknskjdsk}
\begin{aligned}
B_{\ell_{1} \ell_{2}}^{L}(\Delta)=& \frac{1}{\sqrt{\left(2 \ell_{1}+1\right)\left(2 \ell_{2}+1\right)}}\,\left(\begin{array}{lll}
\ell_{1} & \ell_{2} & L \\
0 & 0 & 0
\end{array}\right) 
%\\& \times 
\sum_{M} X_{\ell_{1} \ell_{2}}^{L M *}\left(\hat{\boldsymbol{r}}_{1}, \hat{\boldsymbol{r}}_{2}\right) Y_{L M}(\hat{\boldsymbol{r}})\,.
\end{aligned}
\end{align}
Here $\xi^{(n)}_L$ are given by \eqref{12} and
\begin{align}
\begin{aligned}
X_{\ell_{1}\ell_{2}}^{L M}\left(\hat{\boldsymbol{r}}_{1}, \hat{\boldsymbol{r}}_{2}\right)=&(-1)^{\ell_{1}-\ell_{2}-M} \sqrt{2 L+1} 
%\\& \times 
\sum_{m_{1}, m_{2}}\left(\begin{array}{ccc}
\ell_{1} & \ell_{2} & L \\
m_{1} & m_{2} & -M
\end{array}\right) Y_{\ell_{1} m_{1}}\left(\hat{\boldsymbol{r}}_{1}\right) Y_{\ell_{2} m_{2}}\left(\hat{\boldsymbol{r}}_{2}\right) .
\end{aligned}
\end{align}
%and   
%\bea
%\xi^{(n)}_\ell(r_{12}) &\equiv & \int {\ud k \over 2\pi^2}\, k^{2-n}j_\ell(kr_{12}) P_{\rm m}(k).\label{12a}
%\eea
\fi
Here $\xi^{(n)}_L$ are given by \eqref{12}.
In order to explicitly evaluate the 2PCF \eqref{8} in terms of angles at the observer, we use spherical coordinates $(\varrho,\vartheta,\varphi)$ with $\bm d=(d,0,0)$ along the $z$-axis, and the triangle in the $y=0$ plane oriented such that $\bm r_1$ points in the negative $x$-direction (see  \autoref{fig1}). This gives 
\bea\label{rtp}
&& \hat{\bm r}_1=(1,\phi,\pi),~~ \hat{\bm r}_2=(1,\theta,0),~~ \hat{\bm r}=(1,\gamma,\pi)\,.
\eea
%This form is simplest for evaluating the spherical harmonic coefficients in \autoref{dsjknskjdsk}. Alternatively in terms of Cartesian coordinates we have,
%\bea
%{\hat{\bm r}_1=(-\sin\phi,0,\cos\phi),~~ \hat{\bm r}_2=(\sin\theta,0,\cos\theta),~~ \hat{\bm r}=(-\sin\gamma,0,\cos\gamma)\,, }
%\eea
We also have the relations,
\bea
{r_1}=r\,\frac{\sin(\gamma+\theta)}{\sin(\theta+\phi)}=d\,\frac{\sin\gamma}{\sin(\gamma-\phi)}\,,~~~
{r_2}=r\,\frac{\sin(\gamma-\phi)}{\sin(\theta+\phi)}=d\,\frac{\sin\gamma}{\sin(\gamma+\theta)}
\,.
%r = \Big[r_1^2+r_2^2-2r_1r_2\cos(\theta+\phi) \Big]^{1/2}.
\eea
Using the angles  $\theta,\phi,\gamma\,(=\cos^{-1}\mu)$ and the separation $r$, the 2PCF can be expanded as
\be
\xi_{{g}}(\bm d,\bm r) =\xi_{{g}}(d,\theta,\phi,\mu,r)= b_1\,b_2\sum_{n,\ell'} c_{n\ell'}(d,\theta,\phi,\mu)\,\xi^{(n)}_{\ell'}(r)\,, 
\ee
where the redshift dependence is implicit.
This expansion follows \cite{Matsubara:1999du}, which corrects typos in \cite{Szalay:1997cc} and generalises  \cite{Szalay:1997cc} to include  galaxy bias, unequal redshifts and redshift evolution. In \cite{Matsubara:1999du},  the angular variables are $\theta+\phi$ and $\gamma_i=\cos^{-1}\hat{\bm r}_i\cdot \hat{\bm r}$, with a modified version of $\xi^{(n)}_\ell$. (A simplified and unified form of the expansions in  \cite{Matsubara:1999du} is given in \cite{Bel:2022iuf}.)
The $c_{n\ell}$ and $\tilde c_{n\ell}$ coefficients are discussed in \autoref{app1}. 
\CC{In the general case of $\theta\neq\phi$ and $\phi\neq0$, we find for the galaxy-galaxy correlations,
\begin{align}
c_{00} &= 1 + \frac{1}{3}(\beta_1+\beta_2) + \frac{1}{15}\beta_1\beta_2\big[2+\cos2(\phi+\theta)\big],
\\    c_{20} &= \frac{1}{3}\beta_1\beta_2\alpha_1\alpha_2 \cos(\theta+\phi),
\\    c_{11} &= \frac{1}{5} \alpha_1\Big\{5\beta_1\cos(\phi-\gamma)+\beta_1\beta_2\big[2\cos(\phi - \gamma) + \cos(2\theta+\phi+\gamma)\big]\Big\}
\notag\\
    &~~- \frac{1}{5}\alpha_2\Big\{5\beta_2\cos(\theta + \gamma)+\beta_1\beta_2\big[2\cos(\theta+\gamma) + \cos(2\phi+\theta-\gamma)\big] \Big\},
\\
    c_{02} &=-\frac{1}{6}\beta_1[3\cos(2\phi-2\gamma)+1]-\frac{1}{6}\beta_2[3\cos(2\gamma+2\theta)+1]
   \notag\\
 &   -\frac{1}{42}\beta_1\beta_2[4 + 9\cos(2\gamma+2\theta) + 9\cos(2\phi-2\gamma) + 2\cos(2\theta+2\phi)]
   ,  
\\    c_{22} &= -\frac{1}{6}\alpha_1\alpha_2\beta_1\beta_2\big[\cos(\theta+\phi)+3\cos(\phi-2\gamma - \theta ) \big],
\\ %   c_{13} &= \frac{1}{40}\beta_1\beta_2\alpha_1 \big[15\sin(\phi-2\theta+\gamma)+2\sin(2\theta+\phi-\gamma)-15\sin(2\theta+\phi-\gamma) -4\sin(\phi+\gamma)
%\notag\\&
%-20 \sin^3\gamma\cos(\phi-2\theta) +5\sin(\phi-2\theta - 3\gamma )+ 5\sin(\phi-2\theta + 3\gamma )\big]
%\notag\\&
%+\frac{1}{40} \beta_1 \beta_2\alpha_2\big[2\sin(2\phi+\theta+\gamma)+15\sin(2\phi-\theta-\gamma)-15\sin(2\phi -\theta+\gamma)- 4\sin(2\phi-\theta-3\gamma)
%\notag\\&
%+10\sin^3\gamma \cos(2\phi-\theta)-5\sin(2\phi-\theta + 3\gamma)- 4\sin(\theta -\gamma)\big],
%\\
c_{13} &=\frac{1}{20} \beta_1 \beta_2 \Big\{-\alpha_1\big[\cos (2 \theta+\phi+\gamma)+5 \cos (\phi-3 \gamma-2 \theta)+2 \cos (\phi-\gamma)\big] 
\nonumber\\&~~
+\alpha_2\big[\cos (2 \phi+\theta-\gamma)+5 \cos (2 \phi-3 \gamma-\theta)+2 \cos (\theta+\gamma)\big] \Big\},
\\
 c_{04} &= \frac{1}{280}\beta_1\beta_2\big[6 +35\cos2(\phi - 2\gamma - \theta) +10 \cos2(\phi - \gamma) 
 \notag\\&~~
 + 10 \cos2(\theta+\gamma) + 3\cos2(\phi + \theta)\big].
\end{align}
For the coefficients of the galaxy-magnification 2PCF \eqref{8dop}, we define
\be
\xi_{{\kappa}}(\bm d,\bm r) =\xi_{{\kappa}}(d,\theta,\phi,\mu,r)= b_1f_2\tilde\alpha_2\sum_{n,\ell'} \tilde c_{n\ell'}(d,\theta,\phi,\mu)\,\xi^{(n)}_{\ell'}(r). 
\ee
From this we find:
\begin{align}
    {\tilde c}_{20} &=\frac{1}{3}\alpha_1\beta_1 \cos(\phi+\theta)\,,
    \\
    \tilde c_{11} &= -\frac{1}{5}\beta_1\Big[2\cos(\theta+\gamma) + \cos(2\phi+\theta - \gamma) \Big] - \cos(\theta + \gamma),
    \\
    \tilde c_{22} &= -\frac{1}{6}\alpha_1\beta_1 \Big[\cos(\theta+\phi) + 3\cos(\phi-2\gamma-\theta) \Big],
    \\
    \tilde c_{13} &= \frac{1}{20}\beta_1\Big[\cos(2\phi + \theta -\gamma) + 2\cos(\theta + \gamma) + 5\cos(2\phi -\theta - 3\gamma) \Big].
\end{align}
}
\CC{Using these,}  the power spectrum may be written as
\begin{align}
P_{{g}}(\bm d, \bm k) & = b_1\,b_2\sum_{n=0}\sum_{\ell'=0}\int \ud^3r \,{\rm e}^{-\I\,\k\cdot\r}\,  c_{n\ell'}({d},\theta,\phi,\mu)\,\xi^{(n)}_{\ell'}(r)
\notag \\
&= b_1\,b_2\sum_{n,\ell,\ell'}\,\I^{-\ell} (2\ell+1)
\int \ud r \,r^2\,j_\ell(kr)\, \xi^{(n)}_{\ell'}(r)
\int {\ud\Omega_{{\bm r}}} \,  c_{n\ell'}({d},\theta,\phi,\mu)\,
\cl_\ell(\hat\k\cdot\hat\r)\,,
\end{align}
\CC{with a similar formula for the galaxy-magnification power spectrum.}
%where the summation limit $n=1$ is from the weak field approximation  and the limit $\ell'=4$ is a truncation in multipoles that we impose. 

At this stage we need the series expansion in $r/d$. Writing
\be
b_1b_2c_{n\ell'}({d},\theta,\phi,\mu) = \sum_{p=0}^\infty c_{n\ell'}^{(p)}(\mu)\left(\frac{r}{d}\right)^p\,,
\ee
the coefficients $c_{n\ell'}^{(p)}$ can be expanded in Legendre polynomials. Then comparing with \eqref{djschsbdisuhdishd}, we see that
\be
\Xi^{(p,n)}_{\ell\ell'}(d)=\int\ud\mu\, c_{n\ell'}^{(p)}(\mu)\,\cl_\ell(\mu)\,.
\ee
Therefore 
\be
b_1b_2\,c_{n\ell'}({d},\theta,\phi,\mu) = \sum_{p,\ell''} \Xi^{(p,n)}_{\ell''\ell'}(d)\,\cl_{\ell''}(\mu)\left(\frac{r}{d}\right)^p\,,
\ee
which leads to
\begin{align}\label{sjdkncskdjnscksnss}
P_{{g}}(\bm d, \bm k) &= \sum_{p,n,\ell,\ell',\ell''}\I^{-\ell}\, (2\ell+1)\,
\Xi^{(p,n)}_{\ell''\ell'}
\int \ud r\,r^2 \,j_\ell(kr)\, \xi^{(n)}_{\ell'}(r)\left(\frac{r}{d}\right)^p
\int {\ud\Omega_{{\bm r}}} \, \cl_{\ell''}(\hat{\bm d}\cdot\hat\r)\,
\cl_\ell(\hat\k\cdot\hat\r)
\notag\\
&=4\pi\sum_{p,n,\ell,\ell'}\I^{-\ell}\, 
\Xi^{(p,n)}_{\ell\ell'}\,\cl_\ell(\hat{\bm d}\cdot\hat{\bm k})
\int \ud r \, r^2 \,j_\ell(kr)\, \xi^{(n)}_{\ell'}(r)\left(\frac{r}{d}\right)^p\,.
\end{align}
It follows that
\begin{align}
\mathcal{P}_\ell^{(p)}(k,d) &=4\pi k^p\sum_{n,\ell'}\I^{-\ell}
\Xi^{(p,n)}_{\ell\ell'}
\int r^{2+p}\ud r \,j_\ell(kr)\, \xi^{(n)}_{\ell'}(r)\,,
\end{align}
which recovers \eqref{cppl} and \eqref{pnpll}:
\begin{align}
\mathcal{P}^{(p)}_\ell
&=\frac{2}{\pi\I^{\ell}}\sum_{\ell', n}{\Xi_{\ell\ell'}^{(p,n)}(d)}P^{pn}_{\ell\ell'}(k)\,,
\qquad
P^{pn}_{\ell\ell'}(k)=k^p\int_0^\infty {\ud q}\, q^{2-n}\,P_{\rm m}(q)\, \mathcal{I}^p_{\ell\ell'}(k,q)\,.
\end{align}

In order to compute the Legendre multipoles in $\mu_k$ of the power spectrum, 
\be
\mathcal{P}_\ell(k,d)=\sum_{p=0}^\infty \frac{\mathcal{P}^{(p)}_\ell}{(kd)^p}\,,
\ee
we need to compute:
\begin{enumerate}
\item the coefficients $\Xi_{\ell\ell'}^{(p,n)}$, which are the Legendre multipoles in $\mu$ of the Taylor coefficients of the coefficients of  the 2PCF appearing in \eqref{cppl};
\item the  integrals $\mathcal{I}^p_{\ell\ell'}(k,q)$ as distributions in $k,q$;
\item the power spectrum multipole `weights' $P^{pn}_{\ell\ell'}(k)$.
\end{enumerate}
\CC{For the galaxy-magnification case, the formulas are the same but with 
\be
b_1f_2\tilde\alpha_2 \tilde c_{n\ell'}({d},\theta,\phi,\mu) = \sum_{p=0}^\infty \tilde c_{n\ell'}^{(p)}(\mu)\left(\frac{r}{d}\right)^p\,,
\ee
and the rest of the derivation is the same but with tilde's on relevant variables. }

Before we implement this computation, we briefly discuss the effects of a window function.

\subsection{Window function}

A careful inspection of the terms in the functions $c_{n\ell'}$ indicate that the nonzero $\Xi_{\ell\ell'}^{(p,n)}$ always have $|\ell-\ell'|+p$ as an even number, which as we will see below implies that the distributional integrals $\mathcal{I}^p_{\ell\ell'}(k,q)$ are combinations of delta functions, step functions and derivatives thereof. This in  turn implies that $P^{pn}_{\ell\ell'}(k)$ can be found relatively easily. This is no longer the case when a window function is involved, which we now illustrate (see \cite{Castorina:2017inr,Beutler:2018vpe,Beutler:2021eqq} for more details on window functions). 

We can incorporate a window function $w(\bm r_i)$ into the power spectrum:
\be
\hat P_{{g}}(\bm d, \bm k)=\int \ud^3 \bm r \,{\rm e}^{-\I\,\k\cdot\r}\,W(\bm d,\bm r)\,\xi_{{g}}(\bm d,\bm r)\quad\mbox{where} \quad
W(\bm d,\bm r) = w(\bm r_1)w(\bm r_2)\,.
\ee
Note that the Yamamoto estimator \cite{Yamamoto:2005dz} is then
\be
\big\langle\hat P_L^{{s}}\big\rangle = (2 L+1) \int \frac{\mathrm{d} \Omega_{{\bm k}}}{4 \pi} \int \mathrm{d}^{3}\bm r_{1} \int\,\mathrm{d}^{3}\bm r_{2}\,
\hat P_{{g}}\,\mathcal{L}_L(\hat{\bm k}\cdot\hat{\bm d}).
\ee 
For simplicity we assume azimuthal symmetry for $W$ and expand it as
\be
W(\bm d,\bm r) = \sum_{p',L} W^{(p')}_L(d)\left(\frac{r}{d}\right)^{p'} \mathcal{L}_L(\mu)\,.
\ee
Then in \eqref{sjdkncskdjnscksnss} we use the identity
\be\label{dshjbcsjda}
\cl_{\ell_1}(\mu)\,\cl_{\ell_2}(\mu)=\sum_{\ell}(2\ell+1)
\left(\begin{array}{lll}
\ell_{1} & \ell_{2} & \ell \\
0 & 0 & 0
\end{array}\right)^2  \cl_{\ell}(\mu)\,,
\ee
which leads to
\begin{align}
\hat P_{{g}}(\bm d, \bm k) 
&=4\pi\sum_{p,p',n,\ell,\ell'\ell''L}\I^{-\ell}(2\ell+1) \left(\begin{array}{lll}
\ell & \ell'' & L \\
0 & 0 & 0
\end{array}\right)^2
\Xi^{(p,n)}_{\ell''\ell'}\, W^{(p')}_L(d)\, \cl_\ell(\hat\k\cdot\hat{\bm d})
\notag \\ &~~~\times
\int r^2\ud r \,j_\ell(kr)\, \xi^{(n)}_{\ell'}(r)\left(\frac{r}{d}\right)^{p+p'}\,.
\end{align}
Therefore we have for the wide-angle multipoles of the windowed power spectrum:
\begin{align}\label{dsjbcscskndc}
\hat{\mathcal{P}}^{(p)}_\ell(k,d)
&=\frac{2}{\pi\I^{\ell}}\sum_{p',n,\ell',\ell'',L }
(2\ell+1) \left(\begin{array}{lll}
\ell & \ell'' & L \\
0 & 0 & 0
\end{array}\right)^2
{\Xi_{\ell''\ell'}^{(p,n)}(d)}\,W^{(p')}_L(d)\,P^{n,p+p'}_{\ell\ell'}(k)\,.
\end{align}
Note that we recover the previous results on using
\be
\left(\begin{array}{lll}
\ell & \ell'' & 0 \\
0 & 0 & 0
\end{array}\right) = \frac{(-1)^\ell}{\sqrt{2\ell+1}}\,\delta_{\ell\ell''}\,.
\ee
In \eqref{dsjbcscskndc}, because there is no restriction on $L$, we see that there is no longer a restriction on $|\ell-\ell'|+p$ being even in the resulting integrals.

\subsection{Computation of $\Xi_{\ell\ell'}^{(p,n)}$ \CC{and $\tilde\Xi_{\ell\ell'}^{(p,n)}$}}

In order to compute $\Xi_{\ell\ell'}^{(p,n)}$ and $\tilde\Xi_{\ell\ell'}^{(p,n)}$ we need to compute a series expansion in $x=r/d$ of functions of $r_1,r_2$ and $\theta,\phi$.
From the geometry in \autoref{fig1} we have
\be\label{skjncskjdbvf}
r_1=d \sqrt{t^{2} x^{2}+2 \mu x t+1},~~ r_2=d \sqrt{(t-1)^{2} x^{2}+2 \mu(t-1) x+1}\,,~~x=\frac{r}{d}\,,~~0\le t\le 1\,.
\ee
This implies
\begin{align}
 { \cos\theta }&=\frac{1-(1-t) x \mu}{\sqrt{(1-t)^{2} x^{2}+1-2(1-t) x \mu}}\approx 1-\frac{(1-t)^2}{3}\big[1-{\cl_{2}(\mu)}\big] x^{2}\,, \\
 { \cos\phi }&=\frac{x \mu t+1}{\sqrt{x^{2} t^{2}+2 x \mu t+1}}\approx 1-\frac{t^2}{3}\big[1-{\cl_{2}(\mu)}\big] x^{2}\,.
\end{align}
In the `bisector' case, where  $\theta=\phi$, we have %choosing the root which is regular as $x\to0$
\be
t=\frac{x\mu+y}{2x\mu},~~~y=\sqrt{x^{2} \mu^{2}+1}-1,
\ee
giving\footnote{\CC{Note that truncating the Legendre expansion in $\sin\theta\approx\displaystyle
{\pi}\big[64-40\cl_{2}(\mu)-9\cl_{4}(\mu)+\cdots\big] x/{512}$ means that $\sin^2\theta+\cos^2\theta\neq1$ at each order in $x$. However, in all the expressions for $c_{n\ell}$ only $\cos(m\theta+n\gamma)$, with $m,n$ integers, appears, which means that we only have factors of $\sin\theta\sin\gamma\approx\frac{1}{2}(1-\mu^2)x$, and no truncation of the Legendre series is necessary. }}
\begin{align}
\sin\theta&=\sqrt{\frac{y(1-\mu^{2})}{2 \mu^{2}+y}}\approx\frac{1}{2}\sqrt{1-\mu^2}x,\\ 
\cos\theta&=\sqrt{\frac{\mu^{2}(y+2)}{2 \mu^{2}+y}}\approx 1-\frac{1}{12}\big[1-{\cl_{2}(\mu)}\big] x^{2}.
\end{align}

For the general case (any $t$), we have
\be\label{cjdhsbdjsbdjsbc}
f(r_1)\approx f(d)+f'(d)\cl_1(\mu)\,tx+\frac{1}{6}\Big\{f'(d)+f''(d)-2\big[2f'(d)-f''(d)\big]\cl_2(\mu)\Big\} t^2x^2+\cdots\,,
\ee
where $'=\ud/\ud\ln d$. For a function of $r_2$, we replace $t\to(t-1)$. In the bisector case:
\be
f(r_1)\approx f(d)+\frac{1}{2}f'(d)\cl_1(\mu)x+\frac{1}{24}\left[3f'(d)+f''(d)-2f''(d)\,\cl_2(\mu)\right] \,x^2+\cdots\,,
\ee
and for $r_2$ we replace $x\to -x$. 

Inserting these into the coefficients $c_{n\ell'}(d,\theta,\phi,\mu)$, we expand as a Taylor series. In order to extract the multipoles from products of Legendre polynomials, it is convenient to use \eqref{dshjbcsjda}.

\subsubsection{Hierarchy of terms}

We briefly give an overview of the relative size of the contributions as we include wide-angle effects. 
We are expanding in a series in $x=r/d$, corresponding to  $1/kd$ in the power spectrum, but once wide-angle effects are included like this there are a number of extra contributions which must be consistently included as we have discussed. Once we approach distances with $z\sim1$, $d\sim 1/\cal H$, so  $1/kd\sim {\cal H}/k$ implying the need for the relativistic terms. 
Therefore,
for a fully consistent approach, we need an expansion in powers of $p+n$ where we consider terms 
\be
\left(\frac{r}{d}\right)^p\left(\frac{\cH}{k}\right)^n\sim \left(\frac{1}{kd}\right)^p\left(\frac{\cH}{k}\right)^n\sim \left(\frac{\cH}{k}\right)^{p+n}\,
,
\ee
together.

%{[I dont follow the argument below and it seems unnecessarily complicated. We have a  formula for $f'$ in LCDM (see our 2206.12375):
%\be 
%\frac{f'}{f}=\frac{1}{1+z}\bigg[ 2+f-\frac{3}{2}\Omega_m\Big(1+\frac{1}{f} \Big)\bigg]
%\ee
%We could just show a plot?
%]}

\CC{Furthermore, we include derivatives of the growth rate, biases, and other variables 
{since  these derivatives are typically not negligible.}
{Consider the $\ln d$ derivative of the grwoth rate:}
\begin{align}
x f'(d)=r\frac{{\d}f}{{\d} d}  = {rH_0h(z)} \frac{{\d}f}{{\d}z}\,,
\end{align}
where $h(z)=H(z)/H_0$ {and $d=\int\d z/H$. This can be} compared to
\begin{align}
x f(d)=\frac{r}{d}f(z)  = \frac{rH_0}{\int_0^z {\d}z/h(z)} f(z)\,,
\end{align}
%the pre-factors in these become the same size for $z\sim 1$. Alternatively, we can write
{leading to}
\begin{align}
 \frac{f'(d)}{f(d)}
%= d\frac{{\d}{f}{{\d}d}  
= \left[h(z)\int_0^z\frac{{\d}z}{h(z)}\right] {\frac{f'(z)}{f(z)}}
%\frac{{\d}f}{{\d}z}
\approx z\, {\frac{f'(z)}{f(z)}}\,.
%\frac{{\d}f}{{\d}z}\,,
\end{align}
The term in square brackets is found numerically to be $\approx z$. Therefore for variables whose derivative with respect to redshift is of the order of the variable itself, the derivatives appearing in the wide-angle expansion can be important. 

In the case of the growth rate,
\be 
\frac{{\d}f}{{\d}z}=\frac{f}{1+z}\bigg[ 2+f-\frac{3}{2}\Omega_m\Big(1+\frac{1}{f} \Big)\bigg]\,,
\ee
where the right-hand side is a similar size to $f$~-- for a LCDM model we find the contribution from the derivative of the growth peaks at $\sim$25\% at $z\sim0.5$. Similarly, for a simple bias model $b\sim (1+z)^\sigma$, we have $b'(z)\sim \sigma\,z(1+z)^{\sigma-1}$ so $b'(z)/b=\sigma\,z/(1+z)$. For $\alpha$ it is  more complicated, because the relative importance of $\alpha'$ is strongly dependent on the evolution and magnification biases. As an example, if $b_e=1=\mathcal{Q}$ then the correction peaks at $\sim50\%$ above $z\sim 1$ and can be  more significant than this even at low redshift.  
}

We now discuss the contributions to the most important multipoles in the cases of the line of sight being chosen either as the mid-point ($t=1/2$) or the equal angle bisector ($\phi=\theta$).
We give general formulas for any $t$ in \autoref{kdjsncskcnskjdn}.

\subsubsection{Contributions to galaxy-galaxy multipoles}

\noindent \textbf{\textit{Contributions to the monopole}}\\

\noindent At $O(x^0)$, i.e., the plane-parallel limit, we have
\begin{align}
\Xi^{(0,0)}_{00} &= b_1b_2+ \frac{1}{3}f(b_1+b_2)+\frac{1}{5}f^2\,
\\
\Xi^{(0,2)}_{00} &= \frac{1}{3}\alpha_1\alpha_2f^2 \,.
\end{align}
Note that all terms are evaluated at position $d$, as is the case with all similar formulas below.

At $O(x^1)$ there are only contributions from $\ell'=1$. In the case of the bisector and midpoint ($t=1/2$) geometries:
\begin{align}
\Xi^{(1,1)}_{01} &=\frac{1}{15}f \Big[5(\alpha_1b_2 + \alpha_2b_1) - f (\alpha_1 + \alpha_2) \Big] + \frac{1}{30}f \Big[\alpha'_1(3f + 5b_2) + \alpha'_2(3f+5b_1) \Big] 
\notag \\&
- \frac{1}{6}f\Big(\alpha_1b'_2 + \alpha_2b_1' \Big) + \frac{1}{6}f'(\alpha_1b_2 + \alpha_2b_1),
%\\
%\Xi^{(1,1)}_{03} &=-\frac{15 \pi^{2}}{2^{20}} f^{2}\left(\alpha_{1}+\alpha_{2}\right)\,.
\end{align}
These all have $n=1$, so the overall order of these corrections in the 2PCF is $(\cH/k)(r/d)$ which is equivalent to $(\cH/k)(1/kd)$ in the power spectrum, and they are present even for the case of a single tracer. 
At $O(x^2)\sim O[1/(kd)^2]$, the contributions arise with $n=0$ so have an overall order $(\cH/k)^0(r/d)^2\sim (\cH/k)^0(1/kd)^2$, and are consequently similar in size to the $O(x)$ terms on large scales. In the bisector case we have,
\begin{align}
\Xi^{(2,0)}_{00}&=-\frac{4 }{45}f^{2}-\frac{1}{36}f^{\prime}(b_1'+b_2')
-\frac{1}{12}b_1'b_2'+\frac{1}{360}\left[5(b_1+b_2)+6f\right](f^{\prime \prime}+3f')
\notag \\&
+\frac{1}{72}\left(f+3b_{2}\right)(b_{1}^{\prime \prime}+3b_1')+\frac{1}{72}\left(f+3b_{1}\right)(b_{2}^{\prime \prime}+3b_2')-\frac{1}{60}(f'^2) \,, \\
\Xi^{(2,0)}_{02}&=-\frac{1}{90}f\Big[2f+9(b_1+b_2) \Big] + \frac{1}{15}f (b_1' + b_2' ) - \frac{1}{15}f'(b_1 +b_2) + \frac{1}{45}f'(b_1' +b_2' ) 
\notag\\&
- \frac{1}{90}f (b_1'' +b_2'') - \frac{1}{630}\Big[12f +7(b_1 +b_2)\Big] + \frac{2}{105}(f')^2.
\end{align}
The bisector and midpoint contributions are no longer equal however. For general $t$, all even $\ell'$ contribute, but in the bisector case only $\ell'=4$ is zero. 

In all these cases there is only marginal simplification from specialising to a single tracer. \\

\noindent \textbf{\textit{Contributions to the dipole}} \\

\noindent  First we have the plane-parallel limit, 
\be
\Xi^{(0,1)}_{11} = \frac{1}{5}{f\left[3 f(\alpha_1-\alpha_2)
+5(\alpha_1b_2-\alpha_2b_1)\right]}\,,
\ee
which vanishes for a single tracer.

At $O(x^1)$ the principal corrections to the dipole are only from $\ell'=0$ and $\ell'=2$ (and not from $\ell'=4$):
\begin{align}
\Xi^{(1,0)}_{10} &=
\frac{1}{6}\left[b_1'(f+3b_2)-b_2'(f+3b_1)
%\nonumber\\&
+f'(b_2-b_1)\right],
\\
\Xi^{(1,0)}_{12} &=\frac{2}{5}f(b_1 - b_2) + \frac{2}{15} \Big[ f(b'_2 -b'_1 ) + f'(b_1 - b_2) \Big]
\,.
\end{align}
These expressions are valid for the bisector and midpoint configurations and vanish in the case of a single-tracer survey. This is not the case for any other configurations and a single-tracer survey will have corrections from all even $\ell'$.

At higher order in $x$, they are further suppressed by a factor of $\cH/k$ (i.e., for $p=2$ the non-zero contributions are from $n=1$), and again the bisector and midpoint configurations have different (complicated) contributions~-- which all vanish for a single tracer, but not in the mutli-tracer case.   \\

\noindent \textbf{\textit{Contributions to the quadrupole}} \\

\noindent  The plane-parallel limit is
\be
\Xi^{(0,0)}_{22} = -\frac{2}{21}{f\left[7(b_1+b_2)+6f\right]}\,,
\ee
and the leading $O(x)$ corrections are for both bisector and midpoint,
\begin{align}
\Xi^{(1,1)}_{21} &= \frac{1}{15}f\Big[f(\alpha_1 + \alpha_2) - 5(\alpha_1b_2 + \alpha_2b_1) \Big] + \frac{1}{15}f\Big[\alpha'_1(3f + 5b_2) + \alpha'_2(3f+5b_1) \Big] 
\notag \\&
- \frac{1}{3}f(\alpha_1b'_2 + \alpha_2b'_1 ) + \frac{1}{3}f'(\alpha_1b_2 + \alpha_2b_1),\\
\Xi^{(1,1)}_{23} &=\frac{4}{35}f^2(\alpha_1 + \alpha_2) - \frac{3}{35}f^2(\alpha'_1 + \alpha'_2)\,.
\end{align}
As in the case of the monopole, these corrections arise from the relativistic terms and are non-zero for a single tracer survey also.
However for consistency at this order we need the $O(x^2)$ contributions, which all have $n=0$. For the bisector case,
\begin{align}
\Xi^{(2,0)}_{20}&=+\frac{4 }{45}f^{2}-\frac{1}{18}f^{\prime}(b_1'+b_2')
-\frac{1}{6}b_1'b_2'+  \frac{1}{180}\big[5(b_1+b_2)+6f\big]f^{\prime \prime}+\frac{1}{36}\left(f+3b_{2}\right)b_{1}^{\prime \prime}
\notag \\&
+\frac{1}{36}\left(f+3b_{1}\right)b_{2}^{\prime \prime}-\frac{1}{30}(f')^2\,, \\
\Xi^{(0,2)}_{22}&= -\frac{2}{3}\alpha_1\alpha_2f^2\,, \\
\Xi^{(2,0)}_{22}&= \frac{1}{882}f\Big[106f + 189(b_1 + b_2)\Big] + \frac{1}{252}b'_1 (22f' -9f ) + \frac{1}{252}b'_2(22f' - 9f)
\notag\\&
- \frac{1}{84}f'[12f + 11(b_1 + b_2)] - \frac{11}{252}f(b''_1 + b''_2) - \frac{1}{1764}f''[132f+77(b_1 +b_2)] 
\notag\\&
+\frac{11}{147}(f')^2,\\
\Xi^{(2,0)}_{24}&=\frac{4}{735}\big[-3f^2+2ff''-2(f'^2)\big]\,.
\end{align}
with a similar formula for the mid-point case.

~\\\noindent \textbf{\textit{Contributions to the higher multipoles}} \\

\noindent  The plane-parallel limit for the octupole is
\begin{align}
\Xi^{(0,1)}_{33}=\frac{2}{5} f^{2}\left(\alpha_{2}-\alpha_{1}\right)\,,
\end{align}
which vanishes for a single tracer, while the hexadecapole is always present:
\begin{align}
\Xi^{(0,0)}_{44}=\frac{8}{35} f^{2}\,.
\end{align}
The leading wide-angle corrections are
\begin{align}
\Xi^{(1,0)}_{32}=-\frac{2}{5}f(b_1 - b_2) + \frac{1}{5}f \Big(b'_2 - b'_1 \Big) +\frac{1}{5}f'(b_1 - b_2),
\end{align}
for the octupole and
\begin{align}
%\Xi^{(1,1)}_{41}&=-\frac{1755 \pi^{2} }{2^{24}} f\left[5 (\alpha_{1} b_{2}+ \alpha_{2} b_{1})-f (\alpha_{1}+ \alpha_{2})\right], \\
\Xi^{(1,1)}_{43}&=-\frac{4}{35}f^2(\alpha_1 + \alpha_2) - \frac{4}{35}f^2(\alpha'_1 + \alpha'_2),
\end{align}
for the hexadecapole. As with the other even multipoles, we also need the $O(x^2)$ terms for consistency:
\begin{align}
\Xi^{(2,0)}_{42}&= -\frac{4}{245}f\Big[6f + 7(b_1 + b_2)\Big] + \frac{4}{35}f'(b_1 + b_2) + \frac{2}{35}b'_1(f' - 2f) + \frac{2}{35}b'_2 (f' - 2f) \\
\notag&
- \frac{1}{35}f\Big(b''_1 +b''_2\Big) - \frac{1}{245}f''\Big[12f + 7(b_1 +b_2)\Big] + \frac{12}{245}(f')^2,\\
\Xi^{(2,0)}_{44}&=\frac{4}{245}f^2 + \frac{2}{35}ff'+\frac{78}{2695}\Big[ff'' -(f')^2 \Big]\,,
\\
  \Xi^{(2,0)}_{64}&= \frac{4}{231}\Big[ff'' -(f')^2 \Big]  \,.
\end{align}
As for the other cases, this is given for the bisector line of sight.\\

In summary, for the even multipoles in a symmetric configuration of $\bm d$ and $\bm r$, the leading wide-angle corrections are suppressed in the 2PCF by a factor of $r/d$ (equivalently $1/kd$ in the power spectrum) but {\it also} by a factor of $\cH/k$, as they arise from the relativistic part. Given that $\cH/k\sim r/d$ for large-scale surveys, this implies that the consistent wide-angle correction needs to include the Newtonian $O(x^2)$ contributions. It also implies that the leading wide-angle corrections require the relativistic corrections for a consistent treatment, \CC{in addition to derivative terms}, even in the single-tracer case, as the Newtonian part does not capture the full range of effects. 

However for the odd multipoles, the multi-tracer plane-parallel limit is already $O(\cH/k)$ from the relativistic corrections, while the leading wide-angle correction does not have this suppression factor (arising purely from the Newtonian part) -- which implies that the Newtonian wide-angle corrections will be a similar size to the relativistic plane-parallel part when $\cH/k\sim r/d$.

\subsubsection{Contributions to galaxy-magnification multipoles}

~\\ \textbf{\textit{Contributions to the monopole}}\\

\noindent At $O(x^0)$, i.e., the plane-parallel limit we  have 
\be
\tilde\Xi^{(0,2)}_{00} = \frac{1}{3}\alpha_1\tilde\alpha_2f^2\,.
\ee
The leading wide-angle corrections at $O(x^2)$, for both bisector and mid-point lines of sight, are 
\begin{align}
\tilde\Xi^{(1,1)}_{01}&=
-\frac{1}{15}\tilde{\alpha}_2f(f-5b_1)- \frac{1}{6}\tilde{\alpha}_2b_1'f + \frac{1}{30}\tilde{\alpha}_2'f(3f+5b_1) + \frac{1}{6}\tilde{\alpha}_2b_1f'\,.
\end{align}
Note that this wide-angle correction to the monopole occurs at order $n=1$ compared with the plane parallel limit which has $n=2$. Therefore we can expect these to be a similar size when $(\cH/k)^2\sim (\cH/k)^1 (r/d)^1$. The next contributions at $O(x^2)$ are all at order $n=2$ and are thus sub-dominant. As in the  galaxy-galaxy case, the midpoint and bisector results are different. 
\begin{align}
    \tilde\Xi^{2,2}_{00} &= -\frac{1}{9}\alpha_1\tilde\alpha_{2}f^2 + \frac{1}{24}\alpha_1\tilde\alpha_{2}'f^2 + \frac{1}{72}\alpha_1'f^2 (3\tilde\alpha_{2} - 2\tilde\alpha_{2}' ) + \frac{1}{12}\alpha_1\tilde\alpha_{2}ff' 
    \notag\\
    &+ \frac{1}{72}f^2 (\alpha_1''\tilde\alpha_{2} +\alpha_1\tilde\alpha_{2}'') + \frac{1}{36}\alpha_1\tilde\alpha_{2} \Big[ff'' - (f')^2\Big], 
     \\
\tilde\Xi^{(2,2)}_{02}&= \frac{1}{18}\alpha_1\tilde{\alpha}_2f^2 + \frac{1}{45}\alpha_1'\tilde\alpha_2'f^2 - \frac{1}{90} \Big[\alpha_1\tilde{\alpha}_2''f^2 + \alpha_1''\tilde{\alpha}_2f^2 + 2\alpha_1\tilde{\alpha}_2ff'' -2\alpha_1\alpha_2(f')^2 \Big]\,.
 \end{align}
~\\\noindent \textbf{\textit{Contributions to the dipole}}\\

\noindent The plane-parallel limit is
\be
\tilde\Xi^{(0,1)}_{11}=-\frac{3}{5}\tilde\alpha_{2}f^2 - \tilde{\alpha}_2b_1f\,,
\ee
while the leading wide-angle $O(x)$ correction is
\begin{align}
\tilde\Xi^{(1,2)}_{10}&=\frac{1}{6}f^2(\alpha_1'\tilde\alpha_{2} - \alpha_1\tilde\alpha_{2}')\,,
%=-\frac{5}{4}\,\tilde\Xi^{(1,2)}_{12} \,.
\\ \tilde\Xi^{(1,2)}_{12}&=-\frac{2 }{15}f^{2}\left({\alpha_{1}^{\prime} \tilde\alpha_{2}}-{ \alpha_{1}\tilde\alpha_{2}^{\prime} }\right)\,.
\end{align}
Note that these are further suppressed, beyond the factor $r/d$ compared to the plane-parallel limit, by an extra factor of $\cH/k$.

At $O(x^2)$ we have corrections from $n=1$ terms which can be a similar size to the $O(x)$ corrections~-- again these are different for the mid-point and bisector cases. For the bisector case, we have
\begin{align}
\tilde\Xi^{(2,1)}_{11}&= \frac{1}{100}\tilde{\alpha}_2f(11f+5b_1) - \frac{1}{40}\tilde{\alpha}_2b_1'f + \frac{1}{200}\tilde{\alpha}_2'(30fb_1' - 11f^2 - 45b_1f) 
\notag \\
& + \frac{3}{40}f'\Big[2\tilde{\alpha}_2b_1' - 2\tilde{\alpha}_2'b_1 
- \tilde{\alpha}_2(2f + 3b_1) \Big] -\frac{3}{200}\tilde{\alpha}_2f''(6f+5b_1) 
\notag \\
&- \frac{1}{200}\tilde{\alpha}_2''(9f^2 + 15b_1f) - \frac{3}{40}\tilde{\alpha}_2b_1''f + \frac{9}{100}\tilde{\alpha}_2(f')^2\,, \\
\tilde\Xi^{(2,1)}_{13}&= -\frac{2}{175}\tilde{\alpha}_2f^2 -\frac{4}{175}\tilde{\alpha}_2'f^2 + \frac{3}{350} \Big[\tilde{\alpha}_2''f^2 +2\tilde{\alpha}_2ff'' - 2\tilde{\alpha}_2(f')^2 \Big]\,.
\end{align}

~\\\noindent \textbf{\textit{Contributions to the quadrupole}}\\

\noindent At $O(x^0)$ we have 
\be
\tilde\Xi^{(0,2)}_{22} = \frac{2}{3}\alpha_1\tilde\alpha_2f^2\,.
\ee
The leading wide-angle corrections at $O(x)$, for the bisector and the midpoint case are , 
\begin{align}
\tilde\Xi^{(1,1)}_{21}&= \frac{1}{15}\tilde{\alpha}_2f(f-5b_1) - \frac{1}{3}\tilde{\alpha}_2b_1'f + \frac{1}{15}\tilde{\alpha}_2'f(3f+5b_1) + \frac{1}{3}\tilde{\alpha}_2b_1f'\,,
\\
\tilde\Xi^{(1,1)}_{23}&=\frac{4}{35}\tilde{\alpha}_2f^2 - \frac{3}{35}\tilde{\alpha}_2'f^2\,.
\end{align}
Again, note that this is a similar size to the plane-parallel contribution. The next order $O(x^2)$ is suppressed, having only contributions from $n=2$: 
\begin{align}
\tilde\Xi^{(2,2)}_{20}&= \frac{1}{9}\alpha_1\tilde\alpha_{2}f^2 - \frac{1}{18}\alpha_1'\tilde\alpha_{2}'f^2 +\frac{1}{36} \Big( \alpha_1\tilde\alpha_{2}''f^2 + \alpha_1''\tilde\alpha_{2}f^2 + 2\alpha_1\tilde\alpha_{2}ff'' \Big) - \frac{1}{18}\alpha_1\tilde\alpha_{2}(f')^2\,,
\\  \tilde\Xi^{(2,2)}_{22}&= - \frac{1}{18}\alpha_1\tilde\alpha_{2}f^2 - \frac{1}{12}\alpha_1\tilde\alpha_{2}'f^2 + \frac{1}{252}\alpha_1' (22\tilde\alpha_{2}'f^2 - 21\tilde\alpha_{2}f^2 ) - \frac{1}{6}\alpha_1\tilde\alpha_{2}ff'
    \notag\\
    &- \frac{11}{252} \Big(\alpha_1''\tilde\alpha_{2}f^2 + \alpha_1\tilde\alpha_{2}''f^2 + 2\alpha_1\alpha_2ff'' \Big) + \frac{11}{126}\alpha_1\tilde\alpha_{2}(f')^2 \,.
\end{align}
~\\
%\noindent 
\textbf{\textit{Contributions to the higher multipoles}}\\

\noindent In the plane-parallel limit, the octupole has
\be
\tilde\Xi^{(0,1)}_{33} = \frac{2}{5}\tilde\alpha_2f^2\,,
\ee
and there is no hexadecapole. 
The leading $O(x)$ wide-angle corrections are
\be
\tilde\Xi^{(1,2)}_{32}=-\frac{ 1}{5}f^{2}\left({\alpha_{1}^{\prime} \tilde\alpha_{2}}-{ \alpha_{1}\tilde\alpha_{2}^{\prime} }\right)\,,
\ee
for the octupole. The $O(x^2)$ contributions are a similar size; for the bisector case, 
\begin{align}
%\tilde\Xi^{(2,1)}_{31}&= -\frac{1}{20} - \frac{11}{100}\tilde{\alpha}_2f^2 - \frac{1}{100} \Big[ 2\tilde{\alpha}_2'f^2 - 6\tilde{\alpha}_2(f')^2 + 3\tilde{\alpha}_2''f^2 + 6\tilde{\alpha}_2ff'' \Big],\\
%
\tilde\Xi^{(2,1)}_{31}&= -\frac{1}{100}\tilde{\alpha}_2f(11f+5b_1) - \frac{1}{10}\tilde{\alpha}_2b_1'f + \frac{1}{50}\tilde{\alpha}_2'\Big[5fb_1' - f(f-5b_1) \Big]
\notag \\
&+ \frac{1}{10}f'(\tilde{\alpha}_2b_1' - \tilde{\alpha}_2'b_1 
+ \tilde{\alpha}_2b_1 ) -\frac{1}{100}\tilde{\alpha}_2f''(6f+5b_1) - \frac{1}{100}\tilde{\alpha}_2''(3f^2 + 5b_1f) 
\notag \\
&- \frac{1}{20}\tilde{\alpha}_2b_1''f + \frac{3}{50}\tilde{\alpha}_2(f')^2\,,
\\
\tilde\Xi^{(2,1)}_{53}&= \frac{1}{63}\Tilde{\alpha}_2f^2 + \frac{2}{63}\Big[\Tilde{\alpha}_2'f^2 +\Tilde{\alpha}_2ff'' - \Tilde{\alpha}_2(f')^2 \Big] + \frac{1}{63}\Tilde{\alpha}_2''f^2\,.
\end{align}

Finally we have the wide-angle correction for the hexadecapole:
\begin{align}
\tilde\Xi^{(1,1)}_{43}&= -\frac{4}{35}\tilde{\alpha}_2f^2 - \frac{4}{35}\tilde{\alpha}_2'f^2 \,,
 \\
\tilde\Xi^{(2,2)}_{42}&= \frac{2}{35} \alpha_1\tilde{\alpha}_2(f')^2 - \frac{1}{35} \Big(\alpha_1''\tilde{\alpha}_2f^2 + \alpha_1\tilde{\alpha}_2''f^2 + 2\alpha_1\tilde{\alpha}_2ff'' - 2\alpha_1'\tilde{\alpha}_2'f^2 \Big)   \,.
\end{align}
\section{Moments of the power spectrum}

In this section, we give a detailed discussion on evaluating the integrals involved in $P^{pn}_{\ell\ell'}(k)$. Some of the results below cover well-known results which we generalise by giving new derivations valid for all values of $p,\ell,\ell'$.

\subsection{Evaluating the integrals $\mathcal{I}^p_{\ell\ell'}(k,q)$}

In general the integral \eqref{ipll} is formally divergent for $p\geq 0$, but its relevance for us is as a distribution, so we need to find its finite part. This is because it is integrated against $P_m(q)$, which will result in a convergent result, provided that $P_m(q)$ is sufficiently compact. For example, for $\ell=\ell'$, we have the well-known closure relation:
\be
\mathcal{I}^0_{\ell\ell}(k,q) = \frac{\pi}{2kq}\,\delta(k-q)\,,
\ee 
which has a singular point at $k=q$. 
However, for $\ell\neq\ell'$ the distributions and singular points become more complicated -- and in some cases quite subtle. We give a detailed discussion of some of the subtleties of the distributions in \autoref{appI}, including also a full derivation of the results presented here. We first give formulas for $p=0$, then demonstrate how to calculate the integrals for $p>0$. 
Note that the integral \eqref{ipll} satisfies a simple symmetry:
\be
\mathcal{I}^p_{\ell\ell'}(k,q)=
%=\int_0^\infty \ud r \,r^{2+p}\, j_\ell(kr)\,j_{\ell'}(qr) 
\mathcal{I}^p_{\ell'\ell}(q,k)\,.
\ee

\noindent{$\bm{p=0}$}\\

Consider $\mathcal{I}^0_{\ell\ell'}(k,q)$,  with $\ell\neq\ell'$, following~\cite{1991JMP....32..642M} (derived in a new way in \autoref{appI}). 
For $\ell-\ell'$ even:
\begin{align}\label{dsjkcbnsjcskn}
\mathcal{I}^0_{\ell \ell^{\prime}}\left(k, q\right)
=\left\{
\begin{array}{l}
g_{\ell \ell^{\prime}}\left({k}, {q}\right) \Theta\left({k}-{q}\right)+\dfrac{\pi}{2 {k} {q}}(-1)^{\left(\ell-\ell^{\prime}\right) / 2} \delta\left({k}-{q}\right) ~~~~\ell>\ell' \,,\\~\\
g_{\ell^{\prime} \ell}\left({q}, {k}\right) \Theta\left({q}-{k}\right)+\dfrac{\pi}{2 {k} {q}}(-1)^{\left(\ell^{\prime}-\ell\right) / 2} \delta\left({k}-{q}\right)~~~~~\ell<\ell'\,,
\end{array}
\right.
\end{align}
where $\Theta$ is the unit step function (we do not need it at the origin as this does not affect the distribution~-- but see \autoref{appI} for the case $p=-2$ where we \emph{do} need it). We can rewrite this as
\bea
\mathcal{I}^0_{\ell \ell^{\prime}}\left(k, q\right)
&=& g_{\ell \ell^{\prime}}\left({k}, {q}\right) \Theta\left({k}-{q}\right)\Theta(\ell-\ell')+g_{\ell^{\prime} \ell}\left({q}, {k}\right) \Theta\left({q}-{k}\right)\Theta(\ell'-\ell)
\notag\\ &&{}
+\dfrac{\pi}{2 {k} {q}}(-1)^{\left(\ell-\ell^{\prime}\right) / 2} \delta\left({k}-{q}\right).
\eea
%where
%\begin{align}
%\Theta\left({k}-{q}\right)=\left\{\begin{array}{ll}
%1, & {k}>{q} \\
%0, & {k}<{q}
%\end{array}\right.\,,
%\end{align}
The function $g$, for ${k}< q$,  is given by
\begin{align}
\begin{aligned}
g_{\ell^{\prime} \ell}\left({q}, {k}\right)=& \frac{\pi }{{q}^{ 3}} \left(\frac{k}{q}\right)^{\ell}\frac{\Gamma\left[\left(\ell+\ell^{\prime}+3\right) / 2\right]}{\Gamma\left(\ell+3 / 2\right) \Gamma\left[\left(\ell'-\ell\right) / 2\right]}\, { }_{2} F_{1}\left(\frac{\ell+\ell^{\prime}+3}{2}, \frac{\ell-\ell^{\prime}}{2}+1 ; \ell+\frac{3}{2} ; \frac{{k}^{2}}{{q}^{2}}\right),
\end{aligned}
\end{align}
and $g_{\ell\ell'}(k,q)=0$. For $k>q$ we have 
\begin{align}
\begin{aligned}
g_{\ell \ell^{\prime}}\left({k}, {q}\right)=& \frac{\pi }{{k}^{ 3}} \left(\frac{q}{k}\right)^{\ell'}\frac{\Gamma\left[\left(\ell+\ell^{\prime}+3\right) / 2\right]}{\Gamma\left(\ell^{\prime}+3 / 2\right) \Gamma\left[\left(\ell-\ell'\right) / 2\right]} \,{ }_{2} F_{1}\left(\frac{\ell+\ell^{\prime}+3}{2}, \frac{\ell^{\prime}-\ell}{2}+1 ; \ell^{\prime}+\frac{3}{2} ; \frac{{q}^{2}}{{k}^{2}}\right).
\end{aligned}
\end{align}
For $\ell-\ell'$ odd we have instead,
\be
\mathcal{I}^0_{\ell \ell^{\prime}}\left(k, q\right) = g_{\ell \ell^{\prime}}\left({k}, {q}\right) \Theta\left({k}-{q}\right) + g_{\ell^{\prime} \ell}\left({q}, {k}\right) \Theta\left({q}-{k}\right)\,.
\ee
These functions are not actually that complicated for the small values of $\ell$ that we need, and in general can be written in terms of Legendre functions~\cite{1991JMP....32..642M}. In the case of $\ell-\ell'$ even, the relevant functions are, for $q>k$,
\begin{align}
g_{20}(q,k) = \frac{3\pi}{2q^3}\,,~~
g_{31}(q,k)  = \frac{5\pi k}{2q^4}\,,~~
g_{40}(q,k) = \frac{5 \pi\left(3 q^{2}-7 k^{2}\right)}{4 q^{5}}\,,~~
%\\g_{60}(q,k) &=\frac{21 \pi\left(33 k^{4}-30 k^{2} q^{2}+5 q^{4}\right)}{16 q^{7}}
g_{42}(q,k) = \frac{7 \pi k^{2}}{2 q^{5}}\,,
%\\g_{62}(q,k) &= -\frac{9 \pi k^{2}\left(11 k^{2}-7 q^{2}\right)}{4 q^{7}}
%\\g_{64}(q,k) &= \frac{11 \pi k^{4}}{2 q^{7}}\,,
\end{align}
and so on. The cases for $k>q$ are given  by swapping $k\leftrightarrow q$. Note that $g_{\ell\ell'}(k,q)$ are not required for $\ell<\ell'$. From this we find
\begin{align}
\mathcal{I}^0_{20}\left(k, q\right)&=\frac{3 \pi }{2 k^{3}}\,\Theta(k-q)-\frac{\pi }{2 kq}\,\delta(k-q)\,,\\
\mathcal{I}^0_{3,1}(k, q)&=\dfrac{5 \pi q }{2 k^{4}}\,\Theta(k-q)-\dfrac{\pi }{2 q^{2}}\,\delta(k-q)\,,
\end{align}
and so on. (Formulas are listed in \autoref{appI}.)

For $\ell-\ell'$ odd, once converted to elementary functions, we can combine the step functions to give the integrals as sums over polynomials and factors of $\ln[|q-k|/(k+q)]$ and $1/(q-k)$. These give singular points to be integrated over later. The lowest $\ell\ell'$ integrals are
\begin{align}
\mathcal{I}^0_{10}\left(k, q\right)&=\frac{1}{k(k^{2}-q^{2})}-\frac{1}{2 k^{2} q}\ln \dfrac{|k-q|}{k+q}\,,\\
\mathcal{I}^0_{2,1}(k, q)&=\dfrac{3 q^{2}-k^{2}}{2 k^{2} q(k^2-q^2)} -\dfrac{k^{2}+3 q^{2} }{4 k^{3} q^{2}}\ln \dfrac{|k-q|}{k+q}\,,\\
\mathcal{I}^0_{30}\left(k, q\right)&= \frac{13 k^{2}-15 q^{2}}{2 k^{3}(k^2- q^{2})}+\frac{3(5 q^{2}- k^{2}) }{4 k^{4} q}\ln \frac{|k-q|}{k+q}\,,\\
%\mathcal{I}^0_{50}\left(k, q\right)&=\frac{\left(-15 k^{4}+210 k^{2} q^{2}-315 q^{4}\right) \ln \left(\frac{|q-k|}{q+k}\right)}{16 k^{6} q}+\frac{113 k^{4}-420 k^{2} q^{2}+315 q^{4}}{8 k^{7}-8 k^{5} q^{2}}\,\\
\mathcal{I}^0_{32}\left(k, q\right)&=\frac{\left(3 q^{2}-3k^{2}\right)\left(k^{2}+3 q^{2}\right)}{8 q^{2} k^{3}(k^2-q^2)}-\frac{3\left(k^{4}+2 k^{2} q^{2}+5 q^{4}\right) }{16 k^{4} q^{3}}\ln \frac{|k-q|}{k+q}\,.
%\mathcal{I}^0_{52}\left(k, q\right)&=-\frac{3\left(k^{6}+5 k^{4} q^{2}+35 k^{2} q^{4}-105 q^{6}\right) \ln \left(\frac{-q+k}{q+k}\right)}{32 q^{3} k^{6}}-\frac{3 k^{6}+13 k^{4} q^{2}-315 k^{2} q^{4}+315 q^{6}}{16 k^{5} q^{2}(k-q)(q+k)}\,\\
%\mathcal{I}^0_{54}\left(k, q\right)&=-\frac{5\left(7 k^{8}+12 k^{6} q^{2}+18 k^{4} q^{4}+28 k^{2} q^{6}+63 q^{8}\right) \ln \left(\frac{-q+k}{q+k}\right)}{256 k^{6} q^{5}}-\frac{105 k^{8}+110 k^{6} q^{2}+136 k^{4} q^{4}+210 k^{2} q^{6}-945 q^{8}}{384 k^{5} q^{4}(k-q)(q+k)}
\end{align}
To find the corresponding formulas with $\ell$ and $\ell'$ reversed, switch $q$ and $k$. These are valid for all $q\neq k$. Further formulas are in \autoref{appI}. \\

\noindent{$\bm{p>0}$}\\

We can derive the distributions for $\mathcal{I}^p_{\ell\ell'}(k,q)$ from $\mathcal{I}^0_{\ell\ell'}(k,q)$ by using
\be
\mathcal{I}^p_{\ell\ell'}(k,q) = -\frac{1}{k^{1-\ell}}\frac{\partial}{\partial k}\left[k^{1-\ell}\int_0^\infty \ud r \,r^{1+p}\, j_{\ell-1}(kr)\,j_{\ell'}(qr)\right] = -k^{\ell-1}\frac{\partial}{\partial k}\left[k^{1-\ell}\mathcal{I}^{p-1}_{\ell-1,\ell'}(k,q)\right],
\ee
and
\be\label{hsbdsjdhbcshjdc}
\mathcal{I}^p_{\ell\ell'}(k,q) = \frac{1}{k^{2+\ell}}\frac{\partial}{\partial k}\left[k^{2+\ell}\int_0^\infty \ud r \,r^{1+p}\, j_{\ell+1}(kr)\,j_{\ell'}(qr)\right] = \frac{1}{k^{2+\ell}}\frac{\partial}{\partial k}\left[k^{2+\ell}\mathcal{I}^{p-1}_{\ell+1,\ell'}(k,q)\right],
\ee
to step up the powers of $r$ in the integrand. These are found via the identities 
\begin{align}
j_{\ell}^{\prime}(x)=j_{\ell-1}(x)-\frac{\ell+1}{x} j_{\ell}(x), \quad j_{\ell}^{\prime}(x)=-j_{\ell+1}(x)+\frac{\ell}{x} j_{\ell}(x)\,.
\end{align}
Then we can derive
\begin{align}\label{djsncksdns}
\mathcal{I}^{p+2}_{\ell\ell'}(k,q) = -k^{\ell-1}\frac{\partial}{\partial k}\left[k^{-2\ell}\frac{\partial}{\partial k}k^{\ell+1}\mathcal{I}^{p}_{\ell\ell'}(k,q)\right]%\nonumber\\ &
= \left[-\frac{\partial^2}{\partial k^2}-\frac{2}{k}\frac{\partial}{\partial k}+\frac{\ell(\ell+1)}{k^2}\right]\mathcal{I}^{p}_{\ell\ell'}(k,q),
\end{align}
to step up two powers of $r$ while keeping $\ell\ell'$ the same. The operator in square brackets in the second equality is the spherical Bessel function differential operator.  From these relations we see that if $|\ell-\ell'|+p $ is even, then the resulting distributions will be a mixture of step functions and $\delta$-functions; while if $|\ell-\ell'|+p $ is odd, the resulting integrals will be rational functions, which have poles of order $p+1$ at $k=q$, plus another rational function with a singular point $\sim\ln|k-q|$.

First we  calculate $\mathcal{I}^1_{\ell \ell^{\prime}}\left(k, q\right)$,  starting with 
\be
\mathcal{I}^1_{00}(k,q) =  \frac{1}{k^{2}}\frac{\partial}{\partial k}\left[k^{2}\mathcal{I}^{0}_{10}(k,q)\right]
%\,,
%\ee
%which gives
%\be
%\mathcal{I}^1_{00}(k,q)
=-\frac{2}{(q+k)^{2}(k-q)^{2}}\,.
\ee
Then
\begin{align}
\mathcal{I}^1_{20}(k,q) &=  \frac{1}{k^{4}}\frac{\partial}{\partial k}\left[k^{4}\mathcal{I}^{0}_{30}(k,q)\right]
%\nonumber\\&
=-\frac{3}{2 k^{3} q} \ln \dfrac{|k-q|}{k+q}+\frac{5 k^{2}-3 q^{2}}{k^{2}(q+k)^{2}(k-q)^{2}}\,,
\end{align}
with similar formulas for other values of $\ell-\ell'$ even. Note  that the singularities in these functions at $k=q$ do not get worse than $ \ln {|k-q|}$ and $1/(k-q)^2$. 

For $\ell-\ell'$ odd, these come from $\ell-\ell'$ even with $p=0$, and thus involve $\delta$-functions. For example,
\begin{align}
\mathcal{I}^1_{10}(k,q) & =  \frac{1}{k^{3}}\frac{\partial}{\partial k}\left[k^{3}\mathcal{I}^{0}_{20}(k,q)\right]= -\frac{ \pi}{2 q^{2}}\,\delta'(k-q)\,,
\\
\mathcal{I}^1_{4,1}(k,q) & =-\frac{4  \pi}{q^{3}}\,\delta(k-q)+\frac{\pi }{2 q^{2}}\,\delta'(k-q)+\frac{35 \pi q }{2 k^{5}}\,\Theta(k-q)\,.
\end{align}
We can also derive general formulas from the closure relation,\footnote{Expressions involving derivatives of delta functions can appear different depending on how they are derived. From
\be
\frac{\partial}{\partial k}\left[q^p\delta(k-q)\right]=q^p\delta'(k-q)=\frac{\partial}{\partial k}\left[k^p\delta(k-q)\right]\,,
\ee
we can derive
\begin{align}
q^p\delta'(k-q)&=k^p\delta'(k-q)+p k^{p-1}\delta(k-q)\,,\\
q^p\delta''(k-q)&=k^p\delta''(k-q)+2pk^{p-1}\delta'(k-q)+(p-1)k^{p-2}\delta(k-q)\,,
\end{align}
and so on. This explains the difference in appearance between some of the expressions here and those in~\cite{Reimberg:2015jma}.
} 
\begin{align}
\mathcal{I}^1_{\ell-1,\ell}(k,q) & = \frac{\pi(\ell+1) }{2 k^{3}}\,\delta(k-q)-\frac{\pi }{2 k^{2}}\,\delta'(k-q)  = \frac{\pi}{2}\frac{q^{\ell-1}}{k^{\ell+1}}\,\delta'(k-q)\,, \\
\mathcal{I}^1_{\ell+1,\ell}(k,q)  &= \frac{\pi  \ell}{2 k^{3}}\,\delta(k-q) +\frac{\pi }{2 k^{2}}\,\delta'(k-q)  = -\frac{\pi}{2}\frac{k^{\ell}}{q^{\ell+2}}\,\delta'(k-q)\,.
\end{align}

For $n\geq2$ we just repeat the process. When $\ell=\ell'$ we can use~\eqref{djsncksdns} and the closure relation, giving
\begin{align}
\mathcal{I}^2_{\ell\ell}(k,q)& =\frac{\pi(\ell+2)(\ell-1) }{2 k^{4}}\,\delta(k-q)+\frac{\pi }{k^{3}}\,\delta'(k-q)-\frac{\pi }{2 k^{2}}\,\delta''(k-q)\,,\\
\mathcal{I}^4_{\ell\ell}(k,q)& =\frac{\pi(\ell-1)(\ell-3)(\ell+4)(\ell+2) }{2 k^{6}}\,\delta(k-q)+\frac{4 \pi\left(\ell^{2}+\ell-3\right) }{k^{5}}\,\delta'(k-q)\nonumber\\
&-\frac{\pi(\ell+3)(\ell-2) }{k^{4}}\,\delta''(k-q)-\frac{2 \pi }{k^{3}}\,\delta'''(k-q)+\frac{\pi }{2 k^{2}}\,\delta^{(4)}(k-q)\,.
\end{align}
We can derive general formulas for nearby $\ell$, such as
\begin{align}
\mathcal{I}^2_{\ell-2,\ell}(k,q)& = \frac{\pi \ell(\ell-1) }{2 k^{4}}\,\delta(k-q)+\frac{\pi\ell }{k^{3}}\,\delta'(k-q)+\frac{\pi }{2 k^{2}}\,\delta''(k-q)\,,\\
\mathcal{I}^2_{\ell+2,\ell}(k,q)& = \frac{ \pi(\ell+2)(\ell+1) }{2 k^{4}}\,\delta(k-q)-\frac{\pi(\ell+1) }{k^{3}}\,\delta'(k-q)+\frac{\pi }{2 k^{2}}\,\delta''(k-q)\,,
\end{align}
with similar formulas for $\mathcal{I}^3_{\ell\pm3,\ell}(k,q)$, $\mathcal{I}^3_{\ell\pm1,\ell}(k,q)$, $\mathcal{I}^4_{\ell\pm4,\ell}(k,q)$, $\mathcal{I}^4_{\ell\pm2,\ell}(k,q)$\,.

Tabulated integrals may be found in \autoref{appI}.

\subsection{Integration of the power spectrum}

To complete our expansion of the multipoles of the power spectrum, we need to compute
\be\label{dsbjhdbcsjds}
P^{pn}_{\ell\ell'}(k)=k^p\int_0^\infty {\ud q}\, q^{2-n}\,P_{\rm m}(q)\, \mathcal{I}^p_{\ell\ell'}(k,q)\,.
\ee
Having evaluated $\mathcal{I}^p_{\ell\ell'}(k,q)$ in terms of relatively simple distributions means that these are straightforward to compute numerically, rather than having to compute highly oscillatory triple integrals. Given that they are distributions, there are some subtleties:
\begin{description}
\item[$|\ell-\ell'|+p $ even:] The integrals consist of delta functions and derivatives thereof, together with step functions. The delta functions evaluate  to sample the power spectrum and its derivatives at $k$, and the step functions sample the long wavelength part of the power spectrum for $\ell<\ell'$ and the short wavelength part for $\ell>\ell'$.
\item[$|\ell-\ell'|+p $ odd:] The integrals consist of regular plus singular parts, all of the form
\be
\int_0^\infty \ud q\, \left\{ f_1(q,k)\big[\ln|k-q|-\ln(k+q)\big]+\frac{f_2(q,k)}{(k-q)^{p+1}}\right\}\,,
\ee
which is singular at $k=q$ and formally diverges. However what we need is the finite part~-- i.e. the Cauchy Principal Value in the case of the  $1/(k-q)$ singularities, and Hadamard regularisation for $p>0$. The stronger singularities for $p>0$ can be evaluated integrating by parts, assuming $P_{\rm m}(q)$ and its derivatives vanish sufficiently rapidly at $q=0$ and $q=\infty$: 
\begin{align}
&\int_0^\infty \ud q\, \left\{ f_1(q,k)\big[\ln|k-q|-\ln(k+q)\big]+\frac{f_2(q,k)}{(k-q)^{p+1}}\right\}
\\ 
&=\int_0^\infty \ud q\, \bigg\{- f_1(q,k)\ln(k+q)
\nonumber\\\nonumber &\qquad\qquad\quad~
+(k-q)(1-\ln|k-q|)\left[-\frac{\partial f_1(q,k)}{\partial q}+\frac{(-1)^{p+1}}{p!}\frac{\partial^{p+2} f_2(q,k)}{\partial q^{p+2}}\right]\bigg\}.
\end{align}
In general we would use finite limits, which are most easily dealt with in this formula by cutting off $P(q)$ with step functions~-- these then evaluate to delta functions in the integrand (and cannot be ignored). More details are given in \autoref{dsjkcnsdkcnskc}.

\end{description}

An alternative approach which avoids the singular integrals is to use \eqref{djsncksdns} and \eqref{hsbdsjdhbcshjdc} to reduce the order in \eqref{dsbjhdbcsjds} down to $\mathcal{I}^{-2}_{\ell\ell'}(k,q) $. For example, for $p$ even, 
\begin{align}\label{kjdsnvsdbsjhk}
P^{pn}_{\ell\ell'}(k) = \left[\mathcal{D}^{(\ell)}_k\right]^{(p+2)/2}\int_0^\infty {\ud q}\, q^{2-n}\,P_{\rm m}(q)\, \mathcal{I}^{-2}_{\ell\ell'}(k,q)\,,
\end{align}
where
\be
\mathcal{D}^{(\ell)}_k=-\frac{\partial^2}{\partial k^2}-\frac{2}{k}\frac{\partial}{\partial k}+\frac{\ell(\ell+1)}{k^2}\,.
\ee
For $p$ odd, 
\begin{align}\label{kldmdskvmlds}
P^{pn}_{\ell\ell'}(k) = \left[\mathcal{D}^{(\ell)}_k\right]^{(p+1)/2}
\frac{1}{k^{2+\ell}}\frac{\partial}{\partial k}\left[k^{2+\ell}
\int_0^\infty {\ud q}\, q^{2-n}\,P_{\rm m}(q)\, \mathcal{I}^{-2}_{\ell+1,\ell'}(k,q)\right]\,.
\end{align}
Therefore, given $P^{-2,n}_{\ell\ell'}(k)$, we can compute the rest simply by taking suitable derivatives. 

There is a  subtlety involved in swapping the derivative and integral because the integrals for $p\geq0$ are distributions and not convergent~-- yet the integrals in \eqref{kjdsnvsdbsjhk} and \eqref{kldmdskvmlds} are convergent and well defined. A full discussion of this is given in \autoref{dsjkcnsdkcnskc}.

\subsubsection{Multipoles of the power spectra}

\begin{figure}[t]
\begin{centering}
\includegraphics[width=0.49\textwidth]{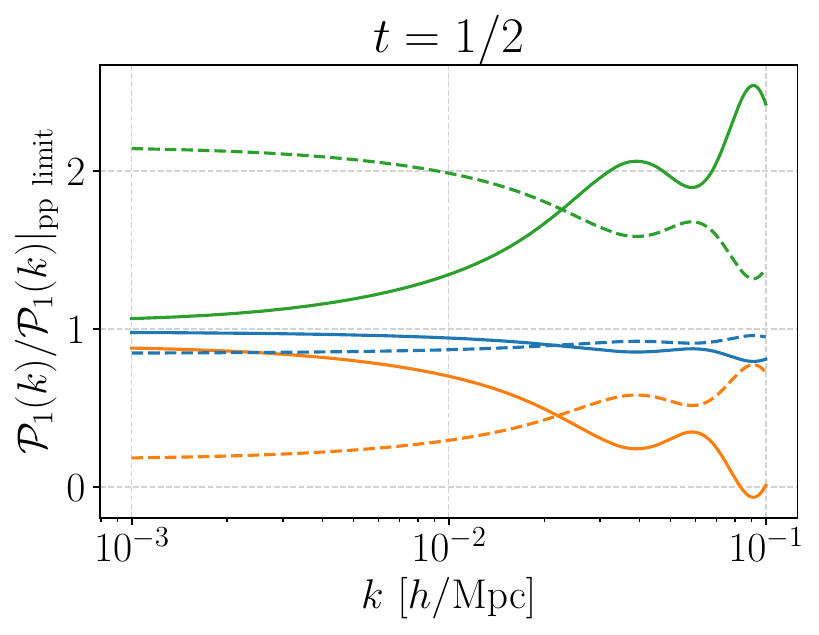}
\includegraphics[width=0.49\textwidth]{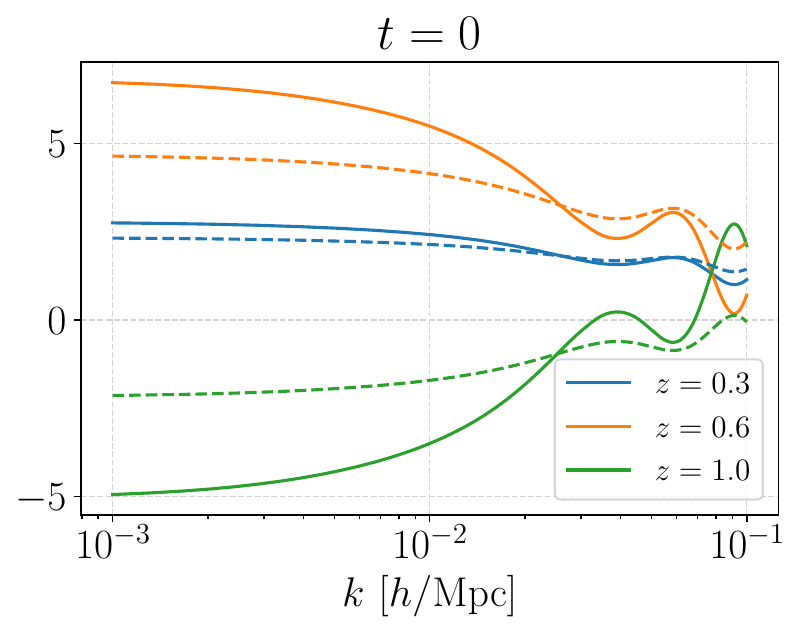}
\caption{Leading-order wide-angle corrections to the dipole of the power spectrum for two tracers, shown relative to the plane-parallel limit at various redshifts. Solid lines are the correct expressions, and dashed correspond to neglecting derivative terms. Two configurations are shown:   bisector ({\em left}) and endpoint ({\em right}). \label{jscndsk}}
\end{centering}
\end{figure}

For completeness, we explicitly give here the lowest multipoles of the power spectra up to $O(1/kd)$. For the galaxy-galaxy bisector case, we have
\begin{align}
\mathcal{P}_0 {(k)}&= \frac{1}{15} P(k)\left[3 f^2+5 f (b_1 + b_2)+15b_1 b_2 \right]+\frac{1}{30 k^2 d}\Big[kP_{,k}(k)+P(k)\Big]\Big[
10f( \alpha_1  b_2+  b_1 \alpha_2)
\nonumber \\
& -2 f^2( \alpha_1 +\alpha_2)
+\alpha_1^{\prime}f (3f+5  b_2)
+\alpha_2^{\prime}f (3f+5  b_1)
\nonumber \\
&-5 f(b_1^{\prime} \alpha_2+ b_2^{\prime} \alpha_1)
%\nonumber \\& 
+5f'( \alpha_1 b_2+ b_1 \alpha_2)  \Big]\,,\\
%%%%%%%%%%%%%%%%%%%
\mathcal{P}_1{(k)}&=
 \frac{{\I} }{{5 k}} P(k)f\left[3 f(\alpha_2 -  \alpha_1)+5 \alpha_1 b_2-5 b_1 \alpha_2\right] \nonumber \\
& +\frac{{\I}}{10kd}\Big\{kP_{,k}(k) \left[(4 f+3f')(b_2-  b_1)-3f(b_2^{\prime} -b_1^{\prime}) +5(b_1^{\prime} b_2-b_2^{\prime}b_1)\right]
\nonumber\\
&+{4P(k)}\left[
(3 f+f')(b_2 - b_1)-f(b_2^{\prime} -b_1^{\prime}) \right]\Big\}\,, \\
%%%%%%%%%%%%%%
\mathcal{P}_2{(k)}&=
 \frac{2}{21}P(k) f\left[6 f+7 (b_1+ b_2)\right] +\frac{1}{21k^2d}\bigg\{ { kP_{,k}(k) }  \Big[-(\alpha_1+\alpha_2) f^2-7f(b_1 \alpha_2+\alpha_1 b_2) 
  \nonumber\\
& +7f'(\alpha_1 b_2+b_1 \alpha_2)
+f(6 f+7b_1)\alpha_2^{\prime}+f(6 f+7b_2)\alpha_1^{\prime}-7f(\alpha_2b_1^{\prime}+\alpha_1b_2^{\prime})\Big] \nonumber \\
& +{14}P ( k ) \Big[-10f^2 (\alpha_1+ \alpha_2) +14f(b_1 \alpha_2+\alpha_1 b_2)-\left(3 f^2+14b_2 f\right)\alpha_1^{\prime} \nonumber\\
&-\left(3 f^2+14f b_1\right)\alpha_2^{\prime}-14(b_1 \alpha_2+\alpha_1 b_2)f^{\prime}+14f(\alpha_2b_1^{\prime}+\alpha_1b_2^{\prime})\Big]\bigg\}\,.
\end{align}
In \autoref{jscndsk} we show a plot of the $O(1/kd)$ wide-angle corrections to the dipole for the multi-tracer bisector case, in order to illustrate the size of the corrections and the importance of the derivative terms. Here we have chosen  $b_1=1,b_{e1}=0=\mathcal{Q}_1$, and $b_2=1+z,~b_{e2}=-5,~\mathcal{Q}_2=2$ for illustration purposes. In the bisector case of $t=1/2$, we see that for small $k$ the wide-angle corrections are small relative to the leading relativistic term, but become important for larger $k$. This is the opposite to the endpoint case of $t=0$, where the smaller $k$ is, the more important the wide-angle corrections are. 

We also see that the inclusion of derivative terms is vital for calculating the wide-angle corrections, since neglecting them gives the wrong values, often by a significant margin. The details of how important they depend on the biases $b_i, b_{ei},{\cal Q}_i$, as well as the triangle configuration that is chosen. For example, when $t=1$ the corrections are much smaller because there are no $b_2'$ factors.  The features from the baryon acoustic oscillations are enhanced from the $kP_{,k}$ contribution which is larger than $P(k)$ for large $k$.

For the case of the galaxy magnification power spectrum we have
\begin{align}
\tilde{\mathcal{P}}_0{(k)}&=
-  \frac{1}{3 k^2}P(k) f^2 \alpha_1 \tilde{\alpha}_2  +\frac{1}{30 k^2 d}\big[kP_{,k}(k) +P(k)\big]\big[2 f \tilde{\alpha}_2(5  b_1-f)+(3 f+5  b_1)f\tilde{\alpha}_2^{\prime}  
\nonumber \\&
 +5\tilde{\alpha}_2(f^{\prime} b_1 - fb_1^{\prime})\big]\,,\\
\tilde{\mathcal{P}}_1{(k)}&=
\frac{{\I }}{5 k} P(k)
f \tilde{\alpha}_2 \left(5 b_1+3 f\right)-\frac{{\I }}{10 k^3 d}\big[3kP_{,k}(k) -2 P(k) \big] f^2\left(\alpha_1\tilde{\alpha}_2^{\prime}-\tilde{\alpha}_2\alpha_1^{\prime}\right)\,,\\
%%%%%%%%%%%%%%%%%%%%%%
\tilde{\mathcal{P}}_2{(k)}&=
 \frac{2}{3 k^2}P(k) f^2 \alpha_1 \tilde{\alpha}_2  +\frac{1}{21k^2d}\bigg\{kP_{,k}(k)\Big[
 -(f^2+7 b_1f+7b_1^{\prime} f-7f^{\prime} b_1) \tilde\alpha_2+\left(6 f+7  b_1\right)f\tilde\alpha_2^{\prime}\Big]
 \nonumber\\
& +P(k) \Big[
2 f \tilde\alpha_2\left(7 b_1-5 f\right)
-f\left(3 f+14 b_1\right)\tilde\alpha_2^{\prime}
+14 \tilde\alpha_2(f^{\prime} b_1-fb_1')
\Big]\bigg\}\,.
\end{align}

\section{Conclusions}

We have given for the first time the wide-angle corrections to the multi-tracer 2PCF and associated power spectrum, including the relativistic Doppler corrections which go beyond the normal redshift space distortion effect. We have also presented the wide-angle corrections for the density-magnification cross-power spectrum. The full-sky power spectrum is expanded as a series in powers of $1/kd$, with each term expanded into Legendre multipoles in the angle between the line-of-sight vector $\bm d$ and the mode vector $\bm k$. The coefficients of this expansion are just the coefficients of the equivalent expansion in $r/d$ of the 2PCF, weighted by an appropriate integral over the matter power spectrum.  We have presented the coefficients for the equal-angle bisector and mid-point cases, as well as for an arbitrary line of sight. 

A key result of our analysis has been to show the importance of the relativistic Doppler corrections \CC{as well as the derivative terms in the expansion -- both typically neglected in previous analysis. While the relativistic terms} enter the observed galaxy number density contrast at $O(\cH/k)$, they only enter the 2PCF at $O[(\cH/k)^2]$ -- except in the multi-tracer case, when the corrections appear at $O(\cH/k)$, and is therefore a useful method for measuring relativistic effects. Similarly the density-magnification 2PCF has also been shown to be an important probe of relativistic effects. However, once we are interested in effects which appear at $O(\cH/k)$, wide-angle effects need to be considered -- since in the 2PCF they arise at $O(r/d)$, and in the power spectrum at $O([1/(kd)]$, which for large-scale surveys is potentially a similar size to $O(\cH/k)$. Therefore, for a fully consistent approach we need an expansion in powers of $p+n$ where we consider terms 
\be
\left(\frac{r}{d}\right)^p\left(\frac{\cH}{k}\right)^n\sim \left(\frac{1}{kd}\right)^p\left(\frac{\cH}{k}\right)^n\sim \left(\frac{\cH}{k}\right)^{p+n}\,
,
\ee
\CC{as similar size, and we need to keep derivative terms at each order for consistency}. In doing this we find, in the galaxy-galaxy case, that the even multipoles at $O(r/d)$ receive relativistic corrections of $O(\cH/k)$, which implies Newtonian wide-angle corrections at $O(x^2)$ are required \CC{for consistency}.  In addition, we find that for the odd multipoles in the multi-tracer case only, Newtonian wide-angle corrections are required for a consistent treatment of the plane-parallel limit. 
\CC{This is clearly seen in Fig.~\ref{jscndsk} where the $O(1/kd)$ terms are a significant correction for large $k$. This also shows the importance of the derivative terms in the wide-angle expansion.} This analysis implies that the relativistic calculations also require $n=2$ potential terms for a fully consistent wide-angle expansion, which is a straightforward extension that we leave for future work.

In the galaxy-magnification cross-power case, we show that as a potential observable of relativistic effects, the dipole and octupole have wide-angle corrections which are suppressed by a factor $(\cH/k)(r/d)$ over the plane-parallel limit. For the even multipoles, the wide-angle corrections are of a similar size to the plane-parallel case.

Finally, we have given a full discussion of the resulting integrals that appear in the wide-angle expansion, i.e., $\mathcal{I}^p_{\ell\ell'}(k,q)$. Although these are well known for some values of $p, \ell, \ell'$, the full set of cases has not been discussed in this context \CC{and given in terms of elementary functions}. Furthermore, in \autoref{appI} we have given a new derivation of these integrals as distributions, together with a discussion of understanding the integrals as the Hadamard finite part, which allows us to give a general formula for an analytic function integrated against a pair of spherical Bessel functions~\eqref{djskndksnskdjc}.

\acknowledgments 
CC is supported by the UK Science \& Technology Facilities Council Consolidated Grant ST/P000592/1.
RM is supported by the South African Radio Astronomy Observatory and the National Research Foundation (Grant No. 75415). 

\clearpage
\appendix

\section{The coefficients 
$c_{n\ell}$ }\label{app1}

The $c_{n\ell}$ for the density-density 2PCF are given by \cite{Matsubara:1999du}, correcting typos in \cite{Szalay:1997cc}. First, we note that
a dimensionless alternative $\breve\xi^{(\breve n)}_\ell$  to our $\xi^{(n)}_\ell$ is used in \cite{Matsubara:1999du} [eq.~(3.45)]: 
\be\label{cnlmtous}
\xi^{(n)}_\ell=(-1)^{\breve n+\ell}\,r^{\ell-2\breve n}\,\breve\xi^{(\breve n)}_\ell \quad \mbox{where}\quad \breve n= {1\over2}(n+\ell)\,.
\ee
{Then \cite{Matsubara:1999du} defines $\breve{c}^{(\breve n)}_\ell$, so that our $c_{n\ell}\,\xi^{(n)}_\ell$ corresponds to $\breve c^{(\breve n)}_\ell\, \breve\xi^{(\breve n)}_\ell$, taking into account the definition of  $\breve\alpha_i$ in \cite{Matsubara:1999du}, which corresponds to our case  via $\alpha_i=r^{\ell-2\breve n} \breve\alpha_i$. 
The triangle configuration
in \cite{Matsubara:1999du} is defined by 3 interior angles: the opening angle $\Theta=\cos^{-1}\hat{\bm r}_1\cdot\hat{\bm r}_2$ and the remaining angles $\gamma_1,\gamma_2$. This configuration
corresponds naturally to an `end-point' line of sight in our set-up: 
\be
\mbox{In \cite{Matsubara:1999du}:}\quad\hat{\bm d}=\hat{\bm r}_1\,,~~ \phi=0\,,~~\Theta=\theta\,,~~ \gamma_1=\gamma\,,~~\gamma_2=\pi-(\gamma+\theta) \,.
\ee
The $\breve{c}^{(\breve n)}_\ell$ given by \cite{Matsubara:1999du} [eqs.~(3.34--42)] become in our notation:  
\bea
c_{00} & =&  1 + \frac{1}{3}(\beta_1+\beta_2) + \frac{1}{15}\beta_1\beta_2\left(2+\cos2\theta\right), \label{cnl} \\
c_{20} & = &  \frac{1}{3}\alpha_1\alpha_2\beta_1\beta_2\cos{\theta},  \\
c_{11} & =&  \alpha_1\beta_1\cos\gamma-\alpha_2\beta_2\cos(\gamma+\theta)
\notag\\&&{}
+\frac{1}{5}\beta_1\beta_2\Big \{ \alpha_1 \big[ 2\cos\gamma+\cos(\gamma+2\theta)\big ] - \alpha_2\big [ \cos(\gamma-\theta) + 2\cos(\gamma+\theta) \big ] \Big \},  \\
c_{02} & = & -\frac{1}{6}\beta_1 \big[1+3\cos(2\gamma) \big] - \frac{1}{6}\beta_2 \big[1+3\cos(2\gamma+2\theta) \big] 
\notag\\&&{}
- \frac{1}{42}\beta_1\beta_2 \big[4 + 9 \cos(2\gamma + 2\theta) + 9 \cos(2\theta) + 2\cos(2\theta) \big],  \\
c_{22} & =  &
-\frac{1}{6} \alpha_1 \alpha_2 \beta_1 \beta_2 \big[ \cos\theta + 3\cos(\theta + 2\gamma) \big],  \\
c_{13} & =&  \frac{1}{20}\beta_1\beta_2\Big \{ \alpha_2 \big[ \cos(\gamma - \theta) + 2\cos(\gamma + \theta) + 5\cos(\theta + 3\gamma) \big]
\notag \\ &&{}
- \alpha_1 \big[ \cos(2\theta + \gamma) + 5 \cos(2\theta + 3\gamma) + 2\cos(\gamma) \big] \Big\}, 
 \\
c_{04} & = & \frac{1}{280}\beta_1\beta_2 \big[ 3\cos(2\theta) + 10\cos(2\gamma) + 10\cos(2\gamma + 2\theta) + 35 \cos(4\gamma + 2\theta) +6 \big] .  
\eea

The bisector line of sight, i.e. $\phi=\theta$ in our notation, corresponds in \cite{Matsubara:1999du} to
$\Theta=2\theta$, $\gamma_1=\gamma-\theta$ and $\gamma_2=\pi-(\gamma+\theta)$. 
The $\breve{c}^{(\breve n)}_\ell$ for the bisector case are  given in \cite{Matsubara:1999du} [eqs.~(3.47--55)], with a different convention for the angle between $\hat{\bm d}$ and $\hat{ \bm r}$, i.e. $\breve\gamma =\pi-\gamma$. This leads to 
%Here we re-arrange the expressions, rewriting $\cos\theta$ in terms of $\sin\theta$ and using \eqref{5} to rewrite $\cos\gamma, \sin\gamma$ in terms of $\mu$. This more clearly separates out the wide-angle contributions which are given by $\sin\theta$, where $\sin\theta=0$ is the plane-parallel (or flat-sky) approximation. It also uses the Legendre mutipoles based on $\gamma$, which recover the standard multipoles in the flat-sky limit. 
\begin{align}
c_{00} &=1+{1\over3}\big(\beta_1+\beta_2\big)+{1\over15}\beta_1\beta_2 \big(2 + \cos4\theta\big),
\label{13}\\
c_{20} & =    \frac{1}{3} \alpha_1\alpha_2\beta_1\beta_2\cos{2\theta},  \\
c_{11} & = \alpha_1\beta_1 \cos(\gamma - \theta) - \alpha_2\beta_2 \cos(\gamma + \theta)
\notag\\&~~
+\frac{1}{5}\beta_1\beta_2 \Big\{ \alpha_1 \big[2\cos(\gamma -\theta) + \cos(\gamma + 3\theta) \big] - \alpha_2\big[ 2\cos(\gamma + \theta) + \cos(\gamma - 3\theta) \big] \Big\},
\label{16}\\
c_{02} &= -\frac{1}{6}\beta_1 \big[1 + 3\cos(2\gamma - 2\theta) \big] - \frac{1}{6}\beta_2(1+ 3\cos(2\gamma + 2\theta) \big] 
\notag\\&~~
- \frac{1}{42} \beta_1\beta_2 \big[ 4 + 2\cos 4\theta + 9\cos(2\gamma - 2\theta) + 9\cos(2\gamma + 2\theta) \big] ,
\label{14}\\
c_{22} &=-\frac{1}{6} \alpha_1\alpha_2\beta_1\beta_2\big[\cos 2\theta + 3\cos 2\gamma \big]
,
\\
c_{13} &= \frac{1}{20} \beta_1 \beta_2 \Big\{ \alpha_2 \big[ \cos(\gamma - 3\theta) + 2\cos(\gamma + \theta) + 5 \cos( 3\gamma - \theta) \big] 
\notag\\&~~
- \alpha_1 \big[ \cos(\gamma + 3\theta) + 2\cos(\gamma - \theta) + 5 \cos(\theta + 3\gamma) \big] \Big\},
\label{17}\\
c_{04} &= \frac{1}{280} \beta_1\beta_2 \big[3 \cos4\theta + 10\cos(2\gamma + 2\theta) + 10\cos(2\gamma - 2\theta) + 35 \cos 4\gamma +6 \big].
\label{15}
\end{align}
%In \eqref{13}--\eqref{17}, the plane parallel limit is given by $\sin\theta=0$. 
}

In the general case of $\theta\neq\phi$ and $\phi\neq0$, we find that
\begin{align}
c_{00} &= 1 + \frac{1}{3}(\beta_1+\beta_2) + \frac{1}{15}\beta_1\beta_2\big[2+\cos2(\phi+\theta)\big]
\\    c_{20} &= \frac{1}{3}\beta_1\beta_2\alpha_1\alpha_2 \cos(\theta+\phi)
\\    c_{11} &= \frac{1}{5} \alpha_1\Big\{5\beta_1\cos(\phi-\gamma)+\beta_1\beta_2\big[2\cos(\phi - \gamma) + \cos(2\theta+\phi+\gamma)\big]\Big\}
\notag\\
    &~~- \frac{1}{5}\alpha_2\Big\{5\beta_2\cos(\theta + \gamma)+\beta_1\beta_2\big[2\cos(\theta+\gamma) + \cos(2\phi+\theta-\gamma)\big] \Big\},
\\
    c_{02} &=-\frac{1}{6}\beta_1[3\cos(2\phi-2\gamma)+1]-\frac{1}{6}\beta_2[3\cos(2\gamma+2\theta)+1]
   \notag\\
 &   -\frac{1}{42}\beta_1\beta_2[4 + 9\cos(2\gamma+2\theta) + 9\cos(2\phi-2\gamma) + 2\cos(2\theta+2\phi)]
   ,  
\\    c_{22} &= -\frac{1}{6}\alpha_1\alpha_2\beta_1\beta_2\big[\cos(\theta+\phi)+3\cos(\phi-2\gamma - \theta ) \big],
\\ %   c_{13} &= \frac{1}{40}\beta_1\beta_2\alpha_1 \big[15\sin(\phi-2\theta+\gamma)+2\sin(2\theta+\phi-\gamma)-15\sin(2\theta+\phi-\gamma) -4\sin(\phi+\gamma)
%\notag\\&
%-20 \sin^3\gamma\cos(\phi-2\theta) +5\sin(\phi-2\theta - 3\gamma )+ 5\sin(\phi-2\theta + 3\gamma )\big]
%\notag\\&
%+\frac{1}{40} \beta_1 \beta_2\alpha_2\big[2\sin(2\phi+\theta+\gamma)+15\sin(2\phi-\theta-\gamma)-15\sin(2\phi -\theta+\gamma)- 4\sin(2\phi-\theta-3\gamma)
%\notag\\&
%+10\sin^3\gamma \cos(2\phi-\theta)-5\sin(2\phi-\theta + 3\gamma)- 4\sin(\theta -\gamma)\big],
%\\
c_{13} &=\frac{1}{20} \beta_1 \beta_2 \Big\{-\alpha_1\big[\cos (2 \theta+\phi+\gamma)+5 \cos (\phi-3 \gamma-2 \theta)+2 \cos (\phi-\gamma)\big] 
\nonumber\\&~~
+\alpha_2\big[\cos (2 \phi+\theta-\gamma)+5 \cos (2 \phi-3 \gamma-\theta)+2 \cos (\theta+\gamma)\big] \Big\},
\\
 c_{04} &= \frac{1}{280}\beta_1\beta_2\big[6 +35\cos2(\phi - 2\gamma - \theta) +10 \cos2(\phi - \gamma) 
 \notag\\&~~
 + 10 \cos2(\theta+\gamma) + 3\cos2(\phi + \theta)\big].
\end{align}

For the coefficients of the galaxy-magnification 2PCF \eqref{8dop}, we obtain: 
\begin{align}
    {\tilde c}_{20} &=\frac{1}{3}\alpha_1\beta_1 \cos(\phi+\theta)\,,
    \\
    \tilde c_{11} &= -\frac{1}{5}\beta_1\Big[2\cos(\theta+\gamma) + \cos(2\phi+\theta - \gamma) \Big] - \cos(\theta + \gamma),
    \\
    \tilde c_{22} &= -\frac{1}{6}\alpha_1\beta_1 \Big[\cos(\theta+\phi) + \cos(\phi-2\gamma-\theta) \Big],
    \\
    \tilde c_{13} &= \frac{1}{20}\beta_1\Big[\cos(2\phi + \theta -\gamma) + 2\cos(\theta + \gamma) + 5\cos(2\phi -\theta - 3\gamma) \Big].
\end{align}

\section{Wide-angle expansion coefficients}\label{kdjsncskcnskjdn}

We collect these in terms of the hierarchy $p+n$, where
\be
\left(\frac{r}{d}\right)^p\left(\frac{\cH}{k}\right)^n\sim \left(\frac{\cH}{k}\right)^{p+n}\,.
\ee
Here we used the fact that for large-scale surveys, the two terms are of a similar order of magnitude.

\subsection{{Coefficients  $\Xi^{(p,n)}_{\ell\ell'}$ in the galaxy-galaxy wide angle expansion for all $t$}}

$\displaystyle \bm{O\left[({\cH}/{k})^0\right]:}$
\begin{align}
  \Xi_{00}^{(0,0)} &= b_1b_2 + \frac{1}{3}(b_1 + b_2)f + \frac{1}{5}f^2\,,
  \\
   \Xi_{22}^{(0,0)} &= -\frac{2}{3}f(b_1+b_2) - \frac{4}{7}f^2\,,
   \\
   \Xi_{44}^{(0,0)} &= \frac{8}{35}f^2 \,.
\end{align}
$\displaystyle \bm{O\left[(r/d)^1\right]\sim  O\left[({\cH}/{k})^1\right]:}$
\begin{align}
     \Xi_{11}^{(0,1)} &= (\alpha_1 b_2 - \alpha_2b_1)f+\frac{3}{5}(\alpha_1 - \alpha_2)f^2,
     \\
      \Xi_{33}^{(0,1)} &= \frac{2}{5}(\alpha_2 - \alpha_1)f^2,
  \\
     \Xi_{10}^{(1,0)} &= \frac{1}{15}\Big\{ f'[3(2t-1)f+5(t-1)b_1+5tb_2] + 5(t-1)(f+3b_1)b_2' + 5t(f+3b_2)b_1'\Big\},
    \\
     \Xi_{12}^{(1,0)} &= -\frac{4}{35}f\big[7t(b_1 +b_2) + 6ft - 3f -7b_1\big]-\frac{4}{15}f\big[b_1't + b_2'(t-1)\big] 
     \nonumber\\
     & + \frac{4}{105}f'\big[7b_1(1-t) +6f(1-2t)-7b_2t\big]\,,
\\ 
     \Xi_{32}^{(1,0)} &=  \frac{4}{35}f\big[  7t(b_1 + b_2)+6ft- 3f -7b_1\big] -\frac{2}{5}f[b_1't+b_2'(t-1)]  \nonumber\\
  & + \frac{2}{35}f'\bigg[6f(1-2t)+7b_1(1-t) - 7b_2t \bigg],
  \\\Xi_{34}^{(1,0)} &=\frac{16}{63}f^2(2t-1) + \frac{32}{315}ff'(2t-1) , 
  \\\Xi_{54}^{(1,0)} &= -\frac{16}{63}f^2(2t-1) +\frac{8}{63}ff'(2t-1).
\end{align}
\noindent $\displaystyle \bm{O\left[({r}/{d})^1\,({\cH}/{k})^1\right]
%\sim O\left[({\cH}/{k})^2\right]
:}$
\begin{align}
    \Xi_{01}^{(1,1)} &=\frac{2}{15}f \Big\{5\alpha_1 b_2t - 5\alpha_2b_1(t-1) 
+ f\big[\alpha_1(3t-2) + \alpha_2(1-3t) \big] \Big\}
\notag\\
&+\frac{1}{3}f\big[ \alpha_{1} b_2'(t-1)- \alpha_{2} b_1't\big] +\frac{1}{15}f'\Big\{3f(\alpha_{1}-\alpha_{2})(2 t-1) -5\big[\alpha_{2} b_1(t-1) -\alpha_{1}b_2t\big]\Big\}
\notag\\
&+\frac{1}{15}f\big[  \alpha_{1}'t(3 f+5 b_2) -\alpha_{2}'(t-1)(3f+5b_1)\big] 
,
\\
\Xi_{21}^{(1,1)} &= -\frac{2}{15}f\Big\{f\big[\alpha_1(3t-2) + \alpha_2(1-3t) \big] 
+ 5 \big[\alpha_1b_2t - \alpha_2b_1(t-1) \big] \Big\}
\notag\\
&+ \frac{2}{3}f\big[\alpha_1b_2'(t-1) - \alpha_2b_1't\big] + \frac{2}{15} f'\big[3f(\alpha_1 - \alpha_2)(2t-1) - 5\alpha_2b_1(t-1) + 5\alpha_1b_2t\big] \notag\\
&+ \frac{2}{15}f \big[\alpha_1't(3f+5b_2) - \alpha_2'(t-1)(3f+5b_1)\big],
\\ 
    \Xi_{23}^{(1,1)} &= - \frac{8}{35}f^2 \Big[3t(\alpha_1 - \alpha_2) - 2\alpha_1 + \alpha_2 \Big]+\frac{6}{35}\Big\{f^2\Big[\alpha'_2(t-1) - \alpha'_1t \Big] - ff'(\alpha_1 - \alpha_2)(2t-1) \Big]\Big\}    ,
  \\
  \Xi_{43}^{(1,1)} &= \frac{8}{35} \Big\{ -f^2\Big[\alpha_1(2-3t) + \alpha_2(3t-1) \Big]+f^2 \Big[\alpha'_2(t-1) - \alpha'_1t \Big] - ff'(\alpha_1 - \alpha_2)(2t-1)  \Big\} . 
\end{align}

\noindent For completeness we also give  $\displaystyle \bm{O\left[({r}/{d})^2\right]:}$
\begin{align}
\Xi_{00}^{(0,2)} &= \frac{1}{3}\alpha_1\alpha_2f^2
\\
   \Xi_{00}^{(2,0)} &= -\frac{4}{45}f^2+\frac{1}{18}b_1'\Big[6t(t-1)b_2'+2 t(t-1)f'+t^{2}(f+3 b_2)\Big]+\frac{1}{18}b_2'\Big\{ 2t(t-1)f' 
    \notag\\
& +\big[(t-1)^{2}(f+3 b_1)\big]\Big\}
   +\frac{1}{15}t\left(t-1\right)f'^2+\frac{1}{90}f'\Big[3f\left( 2t^{2}-2t+1\right) +5b_1(t-1)^{2} +5b_2 t^{2}\Big]
     \notag\\
    &+\frac{1}{18}\Big[t^{2}b_1''(f+3 b_2)
+(t-1)^{2}b_2''(f+3 b_1)\Big] +\frac{1}{90}f''\Big[3\left(2t^{2}-2 t+1\right)f +5(t-1)^{2} b_1+5t^{2}b_2 \Big]
\,,
\\
 \Xi_{20}^{(2,0)} &= \frac{4}{45}f^2+\frac{2}{15}t(t-1)f'^2 + \frac{2}{45}f'\Big[3f(2t-2t^2-1) - 5b_1(t-1)^2 - 5b_2t^2 \Big] 
  \notag\\
&+\frac{2}{9}b_1'\Big[3t(t-1)b_2' + f't(t-1) - t^2(f+3b_2) \Big] + \frac{2}{9}b_2'\Big[f't(t-1) -(f+3b_1)(t-1)^2 \Big] 
 \notag\\
 &+ \frac{1}{9}\big[t^2b_1''(f+3b_2) 
 + (t-1)^2b_2''(f+3b_1)\big] + \frac{1}{45}f''\Big[5b_2t^2 + 5b_1(t-1)^2+3(2t^2 - 2t +1)\Big]
\,,
\\ 
\Xi_{02}^{(2,0)} &= \frac{4}{630}f \Big[f(18t^2 -18t -1) + 21b_1(t-1)^2 + 21b_2t^2 \Big]
+ \frac{8}{105}(f')^2t(1-t)
\notag\\&
+ \frac{2}{315}f' \Big[f(48t - 48t^2 - 6) - 28b_1(t-1)^2 - 28b_2t^2 \Big]
+\frac{4}{45}b'_1\Big[f't(1-t)-ft(2t-3) \Big] 
\notag\\&
+ \frac{4}{45}b'_2\Big[f't(1-t)-f(2t-1)(t-1) \Big]-\frac{2}{45}f\Big[b''_1t^2 +b''_2(t-1)^2 \Big] 
\notag\\&
+ \frac{2}{315}f''\Big[6f(2t - 2t^2 +1) - 7b_1(t-1)^2 - 7b_2t^2 \Big] 
,
\\
\Xi_{22}^{(2,0)} &= \frac{44}{9702}f \Big[f(198t^2 - 198t + 85) + 231b_1(t-1)^2 + 231b_2t^2 \Big]+ \frac{44}{147}(f')^2t(1-t) 
\notag\\&
+ \frac{1}{63}b'_1\Big[22f't(1-t) - ft(11t-12) \Big] + \frac{1}{63}b'_2\Big[22f't(1-t) - f(11t +1)(t-1) \Big] 
\notag\\&
- \frac{11}{63}f\Big[b''_1t^2 + b''_2(t-1)^2 \Big] + \frac{1}{441}f'\Big[f(132t - 132t^2 - 66) - 77b_1(t-1)^2 - 77b_2t^2 \Big]
\notag\\&
+ \frac{1}{441}f''\Big[f(132t - 132t^2 - 30) - 77b_1(t-1)^2 - 77b_2t^2 \Big],
\\
\Xi_{42}^{(2,0)} &=- \frac{192}{1470}f \Big[3f(2t^2 - 2t +1) + 7b_1(t-1)^2 + 7b_2t^2 \Big] + \frac{8}{35}b'_1\Big[f't(1-t) + ft(3t-2) \Big] 
\notag\\&
+ \frac{8}{35}b'_2\Big[f't(1-t) + f(3t-1)(t-1) \Big]  + \frac{24}{245}f'\Big[7b_1(t-1)^2+7b_2t^2 + 4f(3t^2 - 3t +1)\Big] 
\notag\\&
- \frac{4}{35}f \Big[b''_1t^2 + b''_2(t-1)^2 \Big]+ \frac{4}{245}f'' \Big[6f(2t - 2t^2 -1) - 7b_1(t-1)^2 - 7b_2t^2 \Big] 
\notag\\&
 + \frac{48}{245}(f')^2t (1-t)
,
\\
\Xi_{24}^{(2,0)} &= \frac{8}{735}f^2(30t^2 - 30t + 1) + \frac{16}{735}ff'(4t-1)(4t+3) 
+ \frac{32}{735}(f')^2t(t-1) 
\notag\\&
+ \frac{16}{735}ff''(2t^2 - 2t + 1),
\\
\Xi_{44}^{(2,0)} &= -\frac{8}{2695}f^2(390t^2 - 390t + 97) + \frac{4}{2695}ff'(78t^2-78t+19) 
+ \frac{312}{2695}t(t-1)(f')^2 
\notag\\&
+ \frac{156}{2695}ff''(2t^2 - 2t + 1),
\\
\Xi_{64}^{(2,0)} &= \frac{64}{231}f^2(3t^2 - 3t + 1) - \frac{16}{231}ff'(10t^2-10t+3) 
+ \frac{16}{231}t(t-1)(f')^2 
\notag\\&
+ \frac{8}{231}ff''(2t^2 - 2t + 1),
\\
\Xi_{22}^{(0,2)} &= -\frac{2}{3}\alpha_1\alpha_2f^2.
\end{align}

\subsection{Coefficients  $\tilde\Xi^{(p,n)}_{\ell\ell'}$ for the galaxy-magnification wide angle expansion}

$\displaystyle \bm{O\left[({\cH}/{k})^1\right]:}$
\begin{align}
    \tilde\Xi_{11}^{(0,1)} &= -\frac{1}{5}\tilde{\alpha}_2f(3f+5b_1),
    \\
      \tilde\Xi_{33}^{(0,1)} &= \frac{2}{5}\tilde{\alpha}_2f^2 \,.
\end{align}

\noindent $\displaystyle \bm{O\left[({r}/{d})^1\,({\cH}/{k})^1\right]\sim O\left[({\cH}/{k})^2\right]:}$
\begin{align}%\\
    %%%%%%%%%%%%%%%%%%%%%%%%%%
       \tilde\Xi_{00}^{(0,2)} &= \frac{1}{3}\alpha_1\tilde{\alpha}_2f^2 \,, \\
      \tilde\Xi_{22}^{(0,2)} &= -\frac{2}{3}\alpha_1\tilde{\alpha}_2f^2\,, \\
    %%%%%%%%%%%%%%%%%%%%%%%%
     \tilde\Xi_{01}^{(1,1)} &= \frac{2}{15}\tilde{\alpha}_2f^2(1-3t) - \frac{2}{3}\tilde{\alpha}_2b_1f(t-1) - \frac{1}{3}\tilde{\alpha}_2b_1'ft - \frac{1}{15}\tilde{\alpha}_2'f(3f+5b_1)(t-1) 
     \notag\\
     &+ \frac{1}{15}f' \Big[3\tilde{\alpha}_2f(1-2t) - 5\tilde{\alpha}_2b_1(t-1) \Big],
    \\
    \tilde\Xi_{21}^{(1,1)} &= \frac{2}{15}\tilde{\alpha}_2f^2(1-3t) +\frac{2}{3}\tilde{\alpha}_2b_1f(t-1) - \frac{2}{3}\tilde{\alpha}_2b'_1ft - \frac{2}{15}\tilde{\alpha}_2'f(3f+5b_1)(t-1) 
    \notag\\
    &+ \frac{1}{15}f' \Big[6\tilde{\alpha}_2f(1-2t)-10\tilde{\alpha}_2b_1(t-1) \Big],
    %\\
   %\tilde\Xi_{03}^{(1,1)} &= \frac{15\pi^2}{2^{19}}(t-1)\tilde{\alpha}_2f^2 \,,
    \\
    \tilde\Xi_{23}^{(1,1)} &= \frac{8}{35}\tilde{\alpha}_2f^2(3t-1) + \frac{6}{35}\Big[\tilde{\alpha}_2'f^2(t-1)+\tilde{\alpha}_2ff'(2t-1) \Big]\,,
    \\
    \tilde\Xi_{43}^{(1,1)} &= -\frac{8}{35}\tilde{\alpha}_2f^2(3t-1) + \frac{8}{35}\Big[\tilde{\alpha}_2'f^2(t-1) + \tilde{\alpha}_2ff'(2t-1) \Big]\,.
 \end{align}
 
\noindent $\displaystyle \bm{O\left[({r}/{d})^2\,({\cH}/{k})^1\right]\sim O\left[({r}/{d})^1\,({\cH}/{k})^2\right]\sim O\left[({\cH}/{k})^3\right]:}$
\begin{align}%\\   \\
   %%%%%%%%%%%%%%%%%%%%%%%%%%%%%%%%%
\tilde\Xi_{10}^{(1,2)} &= \frac{1}{3}\alpha_1\tilde{\alpha}_2ff{'}(2t-1) + \frac{1}{3}f^2\big[\alpha_1'\tilde{\alpha}_2t + \alpha_1\tilde{\alpha}'_2(t-1)\big], \\
\tilde\Xi_{12}^{(1,2)} &= -\frac{2}{5}\alpha_1\tilde{\alpha}_2f^2(2t-1) - \frac{4}{15}\Big[\alpha_1\tilde{\alpha}_2'f^2(t-1) + \alpha_1'\tilde{\alpha}_2f^2t + \alpha_1\tilde{\alpha}_2ff'(2t-1)\Big],\\
\tilde\Xi_{32}^{(1,2)} &= \frac{2}{5}\alpha_1\tilde{\alpha}_2f^2(2t-1) - \frac{2}{5}\Big[\alpha_1\tilde{\alpha}_2'f^2(t-1) + \alpha_1'\tilde{\alpha}_2f^2t + \alpha_1\tilde{\alpha}_2ff'(2t-1) \Big],
\\
    %%%%%%%%%%%%%%%%%%%%%%%%%%
\tilde\Xi_{11}^{(2,1)} &= \frac{9}{225}\tilde{\alpha}_2f \Big[f(9t^2 - 6t + 5) +15b_1(t-1)^2 \Big] - \frac{1}{10}\tilde{\alpha}_2b_1'ft(3t-4)
- \frac{1}{50}\tilde{\alpha}_2' \Big[30b_1'ft(t-1)
\notag\\
&-f^2(9t-1)(t-1) + 3b_1f(t-1)^2 \Big] - \frac{1}{450}f'\Big\{\tilde{\alpha}_2'\Big[270b_1(t-1)^2 + 162f(t-1)(2t-1)\Big] 
\notag\\
&+ 270\tilde{\alpha}_2b_1't(t-1) + 9\tilde{\alpha}_2\Big[f(18t^2-14t+1) + 15b_1(t-1)^2 \Big] \Big\} - \frac{9}{150}\tilde{\alpha}_2f''\Big[3f(2t^2 - 2t +1)
\notag\\
&+5b_1(t-1)^2 \Big]- \frac{9}{150}\tilde{\alpha}_2''f(3f+5b_1)(t-1)^2 - \frac{3}{10}\tilde{\alpha}_2b_1''ft^2 - \frac{9}{25}\tilde{\alpha}_2(f')^2t(t-1)\,,
   \\
 \tilde\Xi_{31}^{(2,1)} &=  -\frac{9}{225}\tilde{\alpha}_2f\Big[f(9t^2 - 6t +5)+15b_1(t-1)^2 \Big] + \frac{2}{5}\tilde{\alpha}_2b_1'ft(2t-1) + \frac{1}{25}\tilde{\alpha}_2' \Big[-10b_1'ft(t-1)
 \notag\\
 &+ 4f^2(3t-2)(t-1) + 20b_1f(t-1)^2 \Big] + \frac{1}{25}f' \Big\{- \tilde{\alpha}_2'\Big[10b_1(t-1)^2 + 6f(t-1)(2t-1) \Big]
 \notag\\
 &- 10\tilde{\alpha}_2b_1't(t-1) + 2\tilde{\alpha}_2 \Big[f(12t^2 - 11t +4) +10b_1(t-1)^2 \Big] \Big\} - \frac{1}{25}\tilde{\alpha}_2f'' \Big[3f(2t^2 - 2t +1) 
 \notag\\
 &+ 5b_1(t-1)^2 \Big] - \frac{1}{25}\tilde{\alpha}_2''f(3f+5b_1)(t-1)^2 - \frac{1}{5}\tilde{\alpha}_2b_1''ft^2 - \frac{6}{25}\tilde{\alpha}_2(f')^2t(t-1),
 \\
\tilde\Xi_{13}^{(2,1)} &= \frac{2}{175}\tilde{\alpha}_2f^2(24t^2 - 16t -5) + \frac{4}{175}\tilde{\alpha}_2'f^2(9t-1)(t-1) +\frac{4}{175}f'\Big[3\tilde{\alpha}_2'f(2t-1)(t-1) 
\notag\\
&+ \tilde{\alpha}_2f(18t^2-14t+1) \Big] + \frac{6}{175}\Big[\tilde{\alpha}_2ff''(2t^2 - 2t+1) + \tilde{\alpha}_2''f^2(t-1)^2 + 2\tilde{\alpha}_2(f')^2t(t-1) \Big], 
    \\
%RM done up to here
\tilde\Xi_{33}^{(2,1)} &= -\frac{2}{225}\tilde{\alpha}_2f^2(138t^2 - 92t +15) +\frac{1}{225}\tilde{\alpha}_2'f^2(23t-7)(t-1) + \frac{1}{225}f' \Big[46\tilde{\alpha}_2'f(2t-1)(t-1)
\notag\\
&+ \tilde{\alpha}_2f(46t^2-38t+7) \Big] +\frac{23}{225} \Big[\tilde{\alpha}_2ff''(2t^2-2t+1) + \tilde{\alpha}_2''f^2(t-1)^2 +2\tilde{\alpha}_2(f')^2t(t-1)\Big] ,
      \\
         \tilde\Xi_{53}^{(2,1)} &= \frac{4}{63}\tilde{\alpha}_2f^2(15t^2 - 10t +3) - \frac{16}{63}\tilde{\alpha}_2'f^2(2t-1)(t-1) + \frac{8}{63}f' \Big[\tilde{\alpha}_2'f(2t-1)(t-1) 
         \notag\\
         &- \tilde{\alpha}_2f(8t^2 - 7t +2) \Big] + \frac{4}{63} \Big[\tilde{\alpha}_2ff''(2t^2 -2t +1)+\tilde{\alpha}_2''f^2(t-1)^2 + 2\tilde{\alpha}_2(f')^2t(t-1).
 \end{align}
 
 \noindent For completeness we give the other $O(x^2)$ contributions which are\\\\
$\displaystyle \bm{O\left[({r}/{d})^2\,({\cH}/{k})^2\right]\sim O\left[({\cH}/{k})^4\right]\,:}$
\begin{align}%  \\
 \tilde \Xi_{00}^{(2,2)} &= -\frac{1}{9}\alpha_1\tilde{\alpha}_2f^2 + \frac{1}{18}\alpha_1\tilde{\alpha}_2'f^2(t-1)^2 + \frac{1}{18}\alpha_1' \Big[2\tilde{\alpha}_2'f^2t(t-1) 
+ \tilde{\alpha}_2f^2t^2 \Big]
  \notag\\
 &+ \frac{1}{18}f' \Big[2\alpha_1'\tilde{\alpha}_2ft(2t-1)+2\alpha_1\tilde{\alpha}_2'f(2t-1)(t-1)+\alpha_1\tilde{\alpha}_2f(2t^2-2t+1) \Big]
 \notag\\
 & +\frac{1}{18}\Big[\alpha_1\tilde{\alpha}_2ff''(2t^2 - 2t +1) + \alpha_1''\tilde{\alpha}_2f^2t^2 + \alpha_1\tilde{\alpha}_2''f^2(t-1)^2 \Big] +\frac{1}{9}\alpha_1\tilde{\alpha}_2(f')^2 t(t-1) ,
    \\
    \tilde\Xi_{20}^{(2,2)} &= \frac{1}{9}\alpha_1\tilde{\alpha}_2f^2 - \frac{2}{9}\alpha_1\tilde{\alpha}_2'f^2(t-1)^2 + \frac{2}{9}\alpha'_1\Big[\tilde{\alpha}_2'f^2t(t-1) - \tilde{\alpha}_2f^2t^2\Big]
 \notag\\
 &+ \frac{2}{9}f'\Big[\alpha_1'\tilde{\alpha}_2ft(2t-1) + \alpha_1\tilde{\alpha}_2'f(t-1)(2t-1) - \alpha_1\tilde{\alpha}_2f(2t^2-2t + 1)\Big] 
 \notag\\
 &+ \frac{1}{9}\Big[\alpha_1\tilde{\alpha}_2f(f'')(2t^2 -2t +1) + \alpha_1''\tilde{\alpha}_2f^2t^2 + \alpha_1\tilde{\alpha}_2''f^2(t-1)^2 \Big] + \frac{2}{9}\alpha_1\tilde{\alpha}_2(f')^2t(t-1) ,   \\
  \tilde\Xi_{02}^{(2,2)} &=  -\frac{2}{45}\alpha_1\tilde{\alpha}_2f^2(3t^2-3t -2) - \frac{2}{45}\alpha_1\tilde{\alpha}_2'f^2(4t-1)(t-1) 
 - \frac{2}{45}\alpha_1' \Big[2\tilde{\alpha}_2'f^2t(t-1) + \tilde{\alpha}_2f^2t(4t-3) \Big] 
  \notag\\
  &- \frac{1}{45}f' \Big[4\alpha_1'\tilde{\alpha}_2ft(2t-1) + 4\alpha_1\tilde{\alpha}_2'f(t-1)(2t-1) + 2\alpha_1\tilde{\alpha}_2f(8t^2 - 8t+1) \Big] 
  \notag\\
  &- \frac{2}{45} \Big[2\alpha_1\tilde{\alpha}_2ff''(2t^2 - 2t + 1) + 2\alpha_1''\tilde{\alpha}_2f^2t^2 + 2\alpha_1\tilde{\alpha}_2''f^2(t-1)^2 \Big] - \frac{4}{45}\alpha_1\tilde{\alpha}_2(f')^2t(t-1) \Big] ,
   \\
   \tilde\Xi_{22}^{(2,2)} &= \frac{22}{693}\alpha_1\tilde{\alpha}_2f^2(33t^2-33t+8) - \frac{1}{63}\alpha_1\tilde{\alpha}_2'f^2(11t-5)(t-1) - \frac{1}{63}\alpha_1'\Big[22\tilde{\alpha}_2'f^2t(t-1)
   \notag\\
   &+ \tilde{\alpha}_2f^2t(11t-6) \Big] - \frac{1}{63}f' \Big[22\alpha_1'\tilde{\alpha}_2ft(2t-1) + 22\alpha_1\tilde{\alpha}_2'f(2t-1)(t-1) + \alpha_1\tilde{\alpha}_2f(22t^2 - 22t +5) \Big]
   \notag\\
   &- \frac{11}{63}\Big[\alpha_1\tilde{\alpha}_2ff''(2t^2 -2t +1) + \alpha_1''\tilde{\alpha}_2f^2t^2 + \alpha_1\alpha_2''f^2(t-1)^2 \Big] - \frac{22}{63}\alpha_1\tilde{\alpha}_2(f')^2(t-1) ,
   \\
   \tilde\Xi_{42}^{(2,2)} &= -\frac{4}{35}\alpha_1\tilde{\alpha}_2f^2(8t^2 - 8t +3) + \frac{8}{35}\alpha_1\tilde{\alpha}_2'f^2(3t-2)(t-1) + \frac{8}{35}\alpha_1'\Big[\tilde{\alpha}_2f^2t(3t-1) 
   \notag\\
   &- \tilde{\alpha}_2'f^2t(t-1) \Big] + \frac{8}{35}f' \Big[2\alpha_1\tilde{\alpha}_2f(3t^2 - 3t +1) - \alpha_1\tilde{\alpha}_2'f(2t-1)(t-1) - \alpha_1'\tilde{\alpha}_2ft(2t-1) \Big] 
   \notag\\
   &- \frac{4}{35}\Big[\alpha_1\tilde{\alpha}_2ff''(2t^2-2t+1) + \alpha_1''\tilde{\alpha}_2f^2t^2 + \alpha_1\tilde{\alpha}_2''f^2(t-1)^2 \Big] -\frac{8}{35}\alpha_1\alpha_2(f')^2t(t-1) .
\end{align}

\clearpage

\section{Analysis of the integrals $\mathcal{I}^p_{\ell\ell'}(k, q)$}
\label{appI}

\subsection{New derivation of $\mathcal{I}^0_{\ell\ell'}(k, q)$}

Here we give a new derivation of the formulas for $\mathcal{I}^0_{\ell\ell'}(k, q)$, which starts from purely convergent integrals. We begin with 
\be
\mathcal{I}^{-2}_{\ell\ell'}(k,q)=\int_0^\infty \ud r \,j_\ell(kr)\,j_{\ell'}(qr)\,.
\ee
Now, for large $r$,
\be
\lim_{r\to\infty} j_\ell(kr) = \frac{\cos(kr-\pi(\ell+1)/2)}{kr}\,,
\ee
which implies that these integrals will converge absolutely, in contrast to the case with $p=0$, where the integral does not converge. 

First, we define
\begin{align}
\tilde g_{\ell \ell^{\prime}}\!\left({k}, {q}\right)=& \frac{\pi }{4{k}} \left(\frac{q}{k}\right)^{\ell}
\!\frac{\Gamma\left[\left(\ell+\ell^{\prime}+1\right) / 2\right]}{\Gamma\left(\ell+3 / 2\right) \Gamma\left[1-\left(\ell-\ell'\right) / 2\right]}\, 
{ }_{2} F_{1}\!\left(\frac{\ell+\ell^{\prime}+1}{2}, \frac{\ell-\ell^{\prime}}{2}; \ell+\frac{3}{2} ; \frac{{q}^{2}}{{k}^{2}}\right)\,,
\end{align}
which is derived from the result {\sc Maple} gives. However, this misses the full answer for all values of $\ell \ell^{\prime}, k, q$. For $\ell-\ell'$ an odd number or zero, we have
\be
\mathcal{I}^{-2}_{\ell\ell'}(k,q) = \Theta(k-q)\,\tilde g_{\ell' \ell}\left({k}, {q}\right)+\Theta(q-k)\,\tilde g_{\ell \ell'}\left(q,k\right)\,,
\ee
while for $\ell-\ell'$ even, we have
\be
\mathcal{I}^{-2}_{\ell\ell'}(k,q) = \Theta(\ell-\ell')\Theta(k-q)\tilde g_{\ell' \ell}\left({k}, {q}\right) +\Theta(\ell'-\ell)\Theta(q-k)\tilde g_{\ell \ell'}\left(q,k\right)\,.
\ee
These formulas work for $k=q$ provided that we use the definition of the step function:
\begin{align}
\Theta\left({k}-{q}\right)=\left\{\begin{array}{lll}
1 &\quad\mbox{for}\quad & {k}>{q}\,, \\
{1}/{2} &\quad\mbox{for}\quad  & k=q\,, \\
0 &\quad\mbox{for}\quad  & {k}<{q}\,.
\end{array}\right.
\end{align}
The formulas can be converted to elementary functions; for example
\begin{align}
\mathcal{I}^{-2}_{0,0}(k, q)&=\frac{ \pi}{2 q}\Theta(q-k)+\frac{ \pi}{2 k}\Theta(k-q)\,, \\
\mathcal{I}^{-2}_{1,0}(k, q)&=\frac{(k-q)(q+k) }{4 q k^{2}}\ln\dfrac{q+k}{|k-q|}+\frac{1}{2 k}\,, \\
\mathcal{I}^{-2}_{1,1}(k, q)&=\frac{ k \pi}{6 q^{2}}\Theta(q-k)+\frac{ q \pi}{6 k^{2}}\Theta(k-q)\,, \\
\mathcal{I}^{-2}_{2,0}(k, q)&=\frac{\pi(k-q)(q+k) }{4 k^{3}}\Theta(k-q) \,,\\
\mathcal{I}^{-2}_{2,1}(k, q)&=\frac{(k-q)(q+k)\left(k^{2}+3 q^{2}\right) }{16 k^{3} q^{2}}\ln \dfrac{q+k}{|k-q|}-\frac{k^{2}-3 q^{2}}{8 q k^{2}}\,, \\
\mathcal{I}^{-2}_{2,2}(k, q)&=\frac{ k^{2} \pi}{10 q^{3}}\Theta(q-k)+\frac{ q^{2} \pi}{10 k^{3}}\Theta(k-q)\,.
\end{align}
These are similar in form to $\mathcal{I}^{0}_{\ell\ell'}(k,q) $, but without the $1/(k-q)$ singular points. The points where $\ln|k-q|$ causes problems in $\mathcal{I}^{0}_{\ell\ell'}(k,q) $  always appear as $(k-q)\ln|k-q|$ here, which is well behaved as $k\to q$. So these integrals always converge to finite values.

From these well-behaved non-singular formulas we can  derive all $\mathcal{I}^{p}_{\ell\ell'}(k,q) $ for $p\geq-1$, using the differentiation formulas given in the text. In terms of hypergeometric functions, these are not particularly helpful,  although they are easy to derive (but painful to simplify!). However, for low values of $\ell,\ell'$ that we are interested in,  it is straightforward in terms of elementary functions: 
\begin{description}
\item[$\bm{p=-1}$] A single derivative of the $\mathcal{I}^{-2}_{\ell\ell'}(k,q) $ formulas leaves weak singular points $\sim\ln|k-q|$. However, no delta functions appear because in $\mathcal{I}^{-2}_{\ell\ell'}(k,q) $ the step functions are either symmetric in $k$ and $q$ (so that they cancel), or where they appear alone, they are accompanied by a factor $k-q$, so that any delta function appears as $(k-q)\delta(k-q)=0$. Step functions appear without being multiplied by $k-q$. We find for the first few:
\begin{align}
\mathcal{I}^{-1}_{0,0}(k, q)&=\frac{1}{2 k q}\ln \frac{k+q}{|k-q|},\\
\mathcal{I}^{-1}_{1,0}(k, q)&=\frac{\pi }{2 k^{2}}\Theta(k-q),\\
\mathcal{I}^{-1}_{1,1}(k, q)&=-\frac{\left(k^{2}+q^{2}\right) }{4 k^{2} q^{2}} \ln \frac{|k-q|}{k+q}-\frac{1}{2 q k},\\
\mathcal{I}^{-1}_{2,0}(k, q)&=\frac{3q }{4 k^{3}}\ln \frac{|k-q|}{(k+q)}+\frac{1}{4 q k}\ln \frac{k+q}{|k-q|}+\frac{3}{2 k^{2}},\\
\mathcal{I}^{-1}_{2,1}(k, q)&=\frac{q \pi }{2 k^{3}}\Theta(k-q),\\
\mathcal{I}^{-1}_{2,2}(k, q)&=-\frac{\left(3 k^{4}+2 k^{2} q^{2}+3 q^{4}\right) }{16 k^{3} q^{3}}\ln \frac{|k-q|}{k+q}-\frac{3\left(k^{2}+q^{2}\right)}{8 q^{2} k^{2}}.
\end{align}
\item[$\bm{p=0}$] Another derivative implies that  $\ln|k-q|\to 1/(k-q)$ and the lone step functions  now lead to the delta functions given in \eqref{dsjkcbnsjcskn}.  
\end{description}

There are a variety of ways to check that these formulas make sense. For $p=-2, -1$ we can just evaluate them numerically and check the results against these formulas. Alternatively, for low values of $\ell,\ell'$, we can write $\int_{0}^{\infty}  
=\lim_{t\to\infty}\int_{0}^{t}$, and perform the integral analytically. 

For $p=0$ these integrals are divergent, but we can check numerically that the formulas make sense for $q\neq k$. We give examples of  simple cases, starting with $(\ell,\ell',k,q)=1, 1, 2, 3$, which is just an example of the closure relation -- and thus should give zero. However, this is in fact not straightforward:  
\begin{align}
&\int_{0}^{\infty} j_{1}({2} r) j_{1}\left({3} r\right) r^{2} d r
=\lim_{t\to\infty}\int_{0}^{t} j_{1}({2} r) j_{1}\left({3} r\right) r^{2} d r
\\\notag &
=\lim_{t\to\infty}\frac{30 t\sin t+6t \sin (5 t) -5 \cos t+5 \cos (5 t)}{360 t} 
=
\lim_{t\to\infty}\frac{30t \sin t+6t \sin (5 t) +5 \cos (5 t)}{360 t}\,, 
\end{align}
which does not converge. However, the mean of this \emph{is} zero, which is the result of the closure relation.
%It seems that the integral relation implicitly assumes Cesaro summation (?) or some other regularisation thing, which seems to an implicit part of Fourier transforms [according to wikipeadia]. 
Trying other values of $(\ell,\ell',k,q)$ the general result works in the same way. For example, $(2, 3, 2, 3)$ gives numerically 0.07447888703 from the formulas above, so that numerically the integral does not converge as the upper limit $\to\infty$. Rewriting the spherical Bessel functions in terms of sin and cos and integrating gives a limit which oscillates between  
$
-\frac{913}{1440}+\frac{233 \ln 5}{3456}$ and $\frac{163}{288}+\frac{233 \ln 5}{3456},
$
 with a mean value of 0.07447888703. From a distributional point of view, the oscillations cancel out, leaving only the mean.

For $p\geq1$ these results can be checked by integrating against a compact function in $k,q$ to ensure their \emph{distributional form} is correct. We have checked for small values of $|\ell-\ell'|+p $ even, where integrals such as 
\be
\int_0^\infty \ud k\int_0^\infty \ud q \int_0^\infty \ud r \,r^{2+p}\, j_\ell(kr)\,j_{\ell'}(qr)\, k^2 q^2 {\rm e}^{-k^2-q^2}\,,
\ee
can be computed analytically by integrating over $k$ and $q$ first and then computing the $r$ integral. These can then be compared with the same integrals computed with the distributions calculated here, which indeed agree. 
\iffalse
For example, we find for $p,\ell,\ell'$
\begin{align}
0,2,0&=-\frac{\pi^{3 / 2}(7 \sqrt{2}-12 \ln (1+\sqrt{2}))}{32}\\
1,1,0&=\pi/8\\
2,0,0&=\frac{3 \pi^{3 / 2} \sqrt{2}}{32}\\
2,2,0&= \frac{3 \pi^{3 / 2} \sqrt{2}}{32}\\
4,0,0&= \frac{15 \pi^{3 / 2} \sqrt{2}}{32}\\
4,2,0&= \frac{27 \pi^{3 / 2} \sqrt{2}}{32}.
\end{align}
(We couldn't do the same checks for $|\ell-\ell'|+p $ odd analytically.)
\fi

\subsection{Integrating the distributions -- dealing with the singularities}\label{dsjkcnsdkcnskc}

As we saw by calculating everything from $\mathcal{I}^{-2}_{\ell\ell'}(k,q) $, the integrals $\mathcal{I}^{p}_{\ell\ell'}(k,q)$ should give meaningful answers for $p\geq0$ even though the integrals themselves are divergent. One way to see this is to write
\begin{align}
\int_0^\infty {\ud q}\, q^{2-n}\,P_{\rm m}(q) \mathcal{I}^p_{\ell\ell'}(k,q)
&= \left[-\frac{\partial^2}{\partial k^2}-\frac{2}{k}\frac{\partial}{\partial k}+\frac{\ell(\ell+1)}{k^2}\right]^m\int_0^\infty {\ud q}\, q^{2-n}\,P_{\rm m}(q) \mathcal{I}^{p-2m}_{\ell\ell'}(k,q)\,.
\end{align}
While the lhs appears to be divergent, 
by differentiating the integral on the right with $2m=p+2$ or $2m=p+1$ appropriately  we must get a finite answer. Where the integrals give distributions in the form of delta functions it's clear what this means. For the other cases it's a bit more subtle as a brute force numerical or analytical evaluation will give infinite answers (this is a key reason for treating the Fourier transforms as formal mathematical transforms rather than introducing cutoffs in $r$ to try to remain within physical constraints).  We know that all the singular points which are not delta functions come from derivatives of $(k-q)\ln|k-q|$, which give the singular terms of the form
\be
\ln|k-q|,~~~\frac{1}{k-q},~~~\frac{1}{(k-q)^2},\cdots,\frac{1}{(k-q)^{p+1}}\,,
\ee
as we differentiate repeatedly. 
The first is termed weakly singular, the second singular, and the higher powers, hyper-singular points. Consequently,  what we need to understand is 
\be
\frac{\partial^i}{\partial k^i}\int_a^b \ud q f(q) (k-q)\ln|k-q| = \int_a^b \ud q f(q) \frac{\partial^i}{\partial k^i}\left[(k-q)\ln|k-q|\right]\,,
\ee
where we just focus on a region around $q=k$, so $0<a<k<b$. 
On the left we have a regular  expression but on the right we have now apparently created an integral with singular points, which diverges for $i>1$ \cite{MONEGATO2009425}. How do we make sense of this? For $i=1$ we can simplify, 
\begin{align}
&\int_a^b \ud q\, f(q) \ln|k-q|=\int_a^b \ud q\, [f(q)-f(k)]\, \ln|k-q|+\int_a^b \ud q\, [f(k)]\, \ln|k-q|
\\ \nonumber
 &=  %\nonumber\\&
f(k)\left[(b - k)\ln(b - k) + (k - a)\ln(k - a) + a - b\right]+ \int_a^b \ud q\, [f(q)-f(k)]\, \ln|k-q|,
\end{align}
where the remaining integral converges assuming that $f(q)$ is analytic in the neighbourhood of $k=q$. On taking a derivative of this with respect to $k$ we have
\be
\frac{\partial}{\partial k}\int_a^b \ud q\, f(q) \ln|k-q|  = \int_a^b \ud q\, \frac{f(q)-f(k)}{k-q} + f(k)\ln\frac{k-a}{b-k}= \dashint_a^b \ud q\, \frac{f(q)}{k-q}\,,\label{dscskdcscs}
\ee
where $\dashint$ represents the Cauchy Principal Value: 
\be
\dashint_a^b \ud q\, \frac{f(q)}{k-q} = \lim_{\epsilon\to0}\left[\int_a^{k-\epsilon}+\int_{k+\epsilon}^b\right]\ud q\, \frac{f(q)}{k-q}\,.
\ee
That is, a symmetric region about the singular point is removed and the limit taken to zero. Note all the integrals in~\eqref{dscskdcscs} converge and terms involving $f'(k)$ cancel.  We conclude  that when we write an expression like
\be
\frac{\partial}{\partial k}\int_a^b \ud q\, f(q) \ln|k-q|  =\int_a^b \ud q\, \frac{f(q)}{k-q}\,,
\ee
swapping the derivative and integral means that we are only taking the principal value of the integral on the right; instead we should write
\be
\frac{\partial}{\partial k}\int_a^b \ud q\, f(q) \ln|k-q|  =\dashint_a^b \ud q\, \frac{f(q)}{k-q}\,.
\ee 

We can take another derivative~\cite{Zozulya:2015:RDI}:
\begin{align}
\frac{\partial^2}{\partial k^2}&\int_a^b \ud q\, f(q) \ln|k-q|  = \frac{\partial}{\partial k}\left[\int_a^b \ud q\, \frac{f(q)-f(k)}{k-q} + f(k)\ln\frac{k-a}{b-k}\right]\nonumber\\
&=-\int_a^b \ud q\, \frac{f(q)-f(k)-f'(k)(q-k)}{(k-q)^2}+ f(k) \ddashint_a^b \ud q\, \frac{1}{(k-q)^2} + f'(k)\dashint_a^b \ud q\, \frac{f(q)}{k-q} \nonumber\\
&=-\int_a^b \ud q\, \frac{f(q)-f(k)-f'(k)(k-q)}{(k-q)^2}+ f(k)\left[\frac{1}{k-a}+\frac{1}{b-k}\right]+f'(k)\ln\frac{k-a}{b-k}\nonumber\\
&=
 \ddashint_a^b \ud q\, \frac{f(q)}{(k-q)^2}\,.
\end{align}
Here, the notation $\ddashint$ is the Hadamard finite part of the integral, which generalises the Cauchy Principal Value to hyper-singular points. For this we delete a small interval around $k=q$ (it does not have to be symmetric), and take the limit as $\epsilon\to0$, and we ignore any diverging terms.
Alternatively, following \cite{MONEGATO2009425},
\begin{align}
&\frac{\partial^2}{\partial k^2}\int_a^b \ud q\, f(q) \ln|k-q|  = \frac{\partial}{\partial k}\left[\int_a^b \ud q\, \frac{f(q)-f(k)}{k-q} + f(k)\ln\frac{k-a}{b-k}\right]\nonumber\\
&=-\dashint_a^b \ud q\, \frac{f(q)-f(k)}{(k-q)^2}+ f(k) \ddashint_a^b \ud q\, \frac{1}{(k-q)^2} %\nonumber\\&
=-\dashint_a^b \ud q\, \frac{f(q)-f(k)}{(k-q)^2}+ f(k)\left[\frac{1}{k-a}+\frac{1}{b-k}\right]\nonumber\\
&=
 \ddashint_a^b \ud q\, \frac{f(q)}{(k-q)^2}\,.
\end{align}
These expressions differ only by divergent terms which we can miraculously ignore (see \cite{MONEGATO2009425}). 
 In a similar manner we can find the finite part of 
\be
 \ddashint_a^b \ud q\, \frac{1}{(k-q)^{p+1}} = \frac{(-1)^p}{p!}\frac{\partial^p}{\partial k^p}\dashint_a^b \ud q\, \frac{1}{k-q}\,.
\ee

Numerically we find it straightforward to calculate these integrals using integration by parts, which naturally returns the finite part. For example, 
\begin{align}
\dashint_a^b \ud q\, \frac{f(q)}{k-q} & =  \frac{\partial}{\partial k}\int_a^b \ud q\, f(q)\ln|k-q|
%\nonumber\\&
= \int_a^b \ud q\, f(q) \frac{\partial}{\partial k}\ln|k-q|\nonumber\\
& = -\int_a^b \ud q\, f'(q) \ln|k-q|+~\text{boundary terms.}
\end{align}

In general, our integrals consist of regular parts plus singular parts and are of the form
\be
\int_0^\infty \ud q\, \left[ f_1(q,k)[\ln|k-q|-\ln(k+q)]+\frac{f_2(q,k)}{(k-q)^{p+1}}\right]\,.
\ee
The singularities at $k=q$ can be evaluated by parts, assuming that $P_{\rm m}(q)$ and its derivatives vanish sufficiently rapidly at $q=0$ and $q=\infty$. Integrating the last term by parts implies that the non-regular parts of this integral can be written as
\begin{align}
& \int_0^\infty\! \ud q\! \left[ f_1(q,k)\ln|k-q|+\frac{f_2(q,k)}{(k-q)^{p+1}}\right] = \int_0^\infty\!\! \ud q \ln|k-q|\!\left[ f_1(q,k)+\frac{(-1)^{p+1}}{p!}\frac{\partial^{p+1} f_2(q,k)}{\partial q^{p+1}}\right]
\nonumber\\
&= \int_0^\infty \ud q\, (k-q)(1-\ln|k-q|)\left[-\frac{\partial f_1(q,k)}{\partial q}+\frac{(-1)^{p+1}}{p!}\frac{\partial^{p+2} f_2(q,k)}{\partial q^{p+2}}\right],
\end{align}
where we integrated by parts to remove the logarithmic singularity.
Note that the part of the integrand containing  $(k-q)(1-\ln|k-q|)$ is now regular~\cite{Zozulya:2015:RDI}. Therefore we have
\begin{align}
&\int_0^\infty \!\ud q \left\{ f_1(q,k)\Big[\ln|k-q|-\ln(k+q)\Big]+\frac{f_2(q,k)}{(k-q)^{p+1}}\right\}\\ \nonumber
&=\int_0^\infty \!\ud q \left\{- f_1(q,k)\ln(k+q)-(k-q)(1-\ln|k-q|)\left[\frac{\partial f_1(q,k)}{\partial q}-\frac{(-1)^{p+1}}{p!}\frac{\partial^{p+2} f_2(q,k)}{\partial q^{p+2}}\right]\right\}.
\end{align}
In reality we are dealing with an integral over the power spectrum with finite limits so the boundary terms need to be taken into account. In general, %[something still wrong here], 
\begin{align}
&\int_a^b\! \ud q \left[ f_1(q,k)[\ln|k-q|-\ln(k+q)]+\frac{f_2(q,k)}{(k-q)^{p+1}}\right]\nonumber\\
&=\int_a^b\! \ud q \left\{- f_1(q,k)\ln(k+q)+(k-q)(1-\ln|k-q|)\left[-\frac{\partial f_1(q,k)}{\partial q}+\frac{(-1)^{p+1}}{p!}\frac{\partial^{p+2} f_2(q,k)}{\partial q^{p+2}}\right]\right\}\nonumber\\
&+\sum_{i=0}^{p-1}\frac{(p-(i+1))!}{p!(q-k)^{p-i}}\frac{\partial^i f_2(q,k)}{\partial q^i}\bigg|_{q=a}^b +\frac{(-1)^{p+1}}{p!}\ln|k-q|\frac{\partial^{p} f_2(q,k)}{\partial q^{p}}\bigg|_{q=a}^b \nonumber\\
&+ (k-q)(1-\ln|k-q|)\left[- f_1(q,k)+\frac{(-1)^{p}}{p!}\frac{\partial^{p} f_2(q,k)}{\partial q^{p}}\right]\bigg|_{q=a}^b.
\end{align}

\subsection{A general formula for $\displaystyle\int_0^\infty \ud r\, f(r) j_{\ell}(kr)\,j_{\ell'}(qr)$}

Given an analytic function $f(r)$ on $[0,\infty)$ we can derive the general formula: 
\begin{align}
%\displaystyle
\int_0^\infty \ud r\, f(r) j_{\ell}(kr)\,j_{\ell'}(qr)%\nonumber\\
&= \sum_{n=0}^\infty \left[-\frac{\partial^2}{\partial k^2}-\frac{2}{k}\frac{\partial}{\partial k}+\frac{\ell(\ell+1)}{k^2}\right]^{n}
\nonumber\\ &~~~~~~\times
\left[\frac{f^{(2n)}(0)}{(2n)!}\mathcal{I}^{-2}_{\ell\ell'}(k,q)
+\frac{f^{(2n+1)}(0)}{(2n+1)!}\mathcal{I}^{-1}_{\ell\ell'}(k,q)
 \right].\label{djskndksnskdjc}
\end{align}

\section{Integral formulas for $p\geq0$}
\label{app4}

Here we tabulate the lowest order integrals in terms of elementary functions required for calculating multipoles up to $\ell=4$, to order $O(x^2)$.

\begin{align}
\mathcal{I}^0_{1,0}(k, q)&=\dfrac{1}{2 q k^{2}}\ln \dfrac{k+q}{|k-q|}+\dfrac{1}{k(k-q)(k+q)}\,, \\
\mathcal{I}^0_{2,0}(k, q)&=\dfrac{3 \pi }{2 k^{3}}\,\Theta(k-q)-\dfrac{\pi }{2 q^{2}}\,\delta(k-q)\,, \\
\mathcal{I}^0_{2,1}(k, q)&=\dfrac{k^{2}+3 q^{2} }{4 k^{3} q^{2}}\ln \dfrac{k+q}{|k-q|}-\dfrac{k^{2}-3 q^{2}}{2 k^{2} q(k-q)(k+q)}\,, \\
\mathcal{I}^0_{3,0}(k, q)&=\dfrac{3\left(k^{2}-5 q^{2}\right)}{4 k^{4} q} \ln \dfrac{k+q}{|k-q|}+\dfrac{13 k^{2}-15 q^{2}}{2 k^{3}(k-q)(k+q)}\,, \\
\mathcal{I}^0_{3,1}(k, q)&=\dfrac{5 \pi q }{2 k^{4}}\,\Theta(k-q)-\dfrac{\pi }{2 q^{2}}\, \delta(k-q)\,, \\
\mathcal{I}^0_{4,3}(k, q)&=\dfrac{5 k^{6}+9 k^{4} q^{2}+15 k^{2} q^{4}+35 q^{6} }{32 k^{5} q^{4}}\ln \dfrac{k+q}{|k-q|}-\dfrac{15 k^{6}+17 k^{4} q^{2}+25 k^{2} q^{4}-105 q^{6}}{48 k^{4} q^{3}(k-q)(k+q)}\,,\\
\mathcal{I}^1_{0,0}(k, q)&=-\dfrac{2}{(k-q)^{2}(k+q)^{2}}\,, \\
\mathcal{I}^1_{1,0}(k, q)&=-\dfrac{\pi }{2 q^{2}}\,\delta'(k-q)\,, \\
\mathcal{I}^1_{1,1}(k, q)&=\dfrac{1}{2 k^{2} q^{2}}\ln \dfrac{k+q}{|k-q|}-\dfrac{k^{2}+q^{2}}{k q(k+q)^{2}(k-q)^{2}}\,, \\
\mathcal{I}^1_{2,0}(k, q)&=\dfrac{3}{2 q k^{3}}\ln \dfrac{k+q}{|k-q|}+\dfrac{5 k^{2}-3 q^{2}}{k^{2}(k-q)^{2}(k+q)^{2}}\,, \\
\mathcal{I}^1_{2,1}(k, q)&=\dfrac{\pi }{2 q^{3}}\,\delta(k-q)-\dfrac{\pi }{2 q^{2}}\,\delta'(k-q)\,, \\
\mathcal{I}^1_{3,0}(k, q)&=\frac{15 \pi }{2 k^{4}}\,\Theta(k-q)
-\frac{5  \pi}{2 q^{3}}\,\delta(k-q)+\frac{\pi }{2 q^{2}}\, \delta'(k-q)
\,,\\
\mathcal{I}^1_{3,1}(k, q)&=\dfrac{3\left(k^{2}+5 q^{2}\right) }{4 k^{4} q^{2}}\ln \dfrac{k+q}{|k-q|}-\dfrac{3 k^{4}-22 k^{2} q^{2}+15 q^{4}}{2 k^{3} q(k-q)^{2}(k+q)^{2}}, \\
\mathcal{I}^1_{3,2}(k, q)&=\dfrac{\pi }{q^{3}}\,\delta(k-q)-\dfrac{\pi }{2 q^{2}}\,\delta'(k-q)\,,\\
\mathcal{I}^1_{3,3}(k, q)&=\dfrac{3\left(5 k^{4}+6 k^{2} q^{2}+5 q^{4}\right)}{16 k^{4} q^{4}}\ln \dfrac{k+q}{|k-q|}
%\notag\\&
-\dfrac{\left(k^{2}+q^{2}\right)\left(15 k^{4}-22 k^{2} q^{2}+15 q^{4}\right)}{8(k+q)^{2} q^{3} k^{3}(k-q)^{2}}\,, \\
\mathcal{I}^1_{4,0}(k, q)&=\dfrac{15\left(k^{2}-7 q^{2}\right) }{4 k^{5} q}\ln \dfrac{k+q}{|k-q|}+\dfrac{81 k^{4}-190 k^{2} q^{2}+105 q^{4}}{2 k^{4}(k-q)^{2}(k+q)^{2}}, \\
\mathcal{I}^1_{4,1}(k, q)&=\dfrac{35 \pi q }{2 k^{5}}\, \Theta(k-q)-\dfrac{4 \pi }{q^{3}}\,\delta(k-q) +\dfrac{\pi }{2 q^{2}}\,\delta'(k-q)\,, \\
\mathcal{I}^1_{4,2}(k, q)&=\dfrac{3\left(3 k^{4}+10 k^{2} q^{2}+35 q^{4}\right)}{16 q^{3} k^{5}}\ln \dfrac{k+q}{|k-q|}
-\dfrac{9 k^{6}+15 k^{4} q^{2}-145 k^{2} q^{4}+105 q^{6}}{8 k^{4} q^{2}(k-q)^{2}(k+q)^{2}}\,, \\
\mathcal{I}^1_{4,3}(k, q)&=\dfrac{3 \pi }{2 q^{3}}\,\delta(k-q) -\dfrac{\pi }{2 q^{2}}\,\delta'(k-q)\,, \\
\mathcal{I}^1_{4,4}(k, q)&=\dfrac{5\left(k^{2}+q^{2}\right)\left(7 k^{4}+2 k^{2} q^{2}+7 q^{4}\right) }{32 k^{5} q^{5}}\ln \dfrac{k+q}{|k-q|}
\notag\\&
-\dfrac{105 k^{8}-40 k^{6} q^{2}-34 k^{4} q^{4}-40 k^{2} q^{6}+105 q^{8}}{48 q^{4} k^{4}(k-q)^{2}(k+q)^{2}},\\
\mathcal{I}^2_{0,0}(k, q)&=-\dfrac{\pi }{2 q^{2}}\,\delta''(k-q)-\dfrac{ \pi}{q^{3}}\,\delta'(k-q)-\dfrac{\pi }{q^{4}}\,\delta(k-q), \\
\mathcal{I}^2_{1,0}(k, q)&=-\dfrac{8 k}{(k-q)^{3}(k+q)^{3}}, \\
\mathcal{I}^2_{1,1}(k, q)&=-\dfrac{\pi }{2 q^{2}}\,\delta''(k-q)-\dfrac{ \pi}{q^{3}}\,\delta'(k-q), \\
\mathcal{I}^2_{2,0}(k, q)&=\dfrac{\pi }{2 q^{2}}\,\delta''(k-q)-\dfrac{ \pi}{2 q^{3}}\,\delta'(k-q)-\dfrac{\pi }{2 q^{4}}\,\delta(k-q), \\
\mathcal{I}^2_{2,1}(k, q)&=\dfrac{3 }{2 k^{3} q^{2}}\ln \dfrac{k+q}{|k-q|}-\dfrac{\left(3 k^{2}-q^{2}\right)\left(k^{2}+3 q^{2}\right)}{q k^{2}(k-q)^{3}(k+q)^{3}}, \\
\mathcal{I}^2_{2,2}(k, q)&=-\dfrac{\pi }{2 q^{2}}\,\delta''(k-q)-\dfrac{ \pi}{q^{3}}\,\delta'(k-q)+\dfrac{2 \pi }{q^{4}}\,\delta(k-q), \\
\mathcal{I}^2_{3,0}(k, q)&=\dfrac{15 }{2 k^{4} q}\ln \dfrac{k+q}{|k-q|}+\dfrac{33 k^{4}-40 q^{2} k^{2}+15 q^{4}}{k^{3}(k-q)^{3}(k+q)^{3}}, \\
\mathcal{I}^2_{3,1}(k, q)&=\dfrac{\pi }{2 q^{2}}\,\delta''(k-q)-\dfrac{3  \pi}{2 q^{3}}\,\delta'(k-q), \\
\mathcal{I}^2_{3,2}(k, q)&=\dfrac{3\left(3 k^{2}+5 q^{2}\right) }{4 k^{4} q^{3}}\ln \dfrac{k+q}{|k-q|}-\dfrac{9 k^{6}-9 k^{4} q^{2}+31 k^{2} q^{4}-15 q^{6}}{2 k^{3} q^{2}(k-q)^{3}(k+q)^{3}},\\
\mathcal{I}^2_{3,3}(k, q)&=-\dfrac{\pi }{2 q^{2}}\,\delta''(k-q)-\dfrac{ \pi}{q^{3}}\,\delta'(k-q)+\dfrac{5 \pi }{q^{4}}\,\delta(k-q), \\
\mathcal{I}^2_{4,0}(k, q)&=-\dfrac{27 \pi }{2 q^{4}}\,\delta(k-q)+\dfrac{4  \pi}{q^{3}}\,\delta'(k-q)-\dfrac{\pi }{2 q^{2}}\,\delta''(k-q)+\dfrac{105 \pi }{2 k^{5}}\,\Theta(k-q), \\
\mathcal{I}^2_{4,1}(k, q)&=\dfrac{15\left(k^{2}+7 q^{2}\right) }{4 k^{5} q^{2}}\ln \dfrac{k+q}{|k-q|}-\dfrac{15 k^{6}-191 k^{4} q^{2}+265 k^{2} q^{4}-105 q^{6}}{2 q k^{4}(k-q)^{3}(k+q)^{3}}, \\
\mathcal{I}^2_{4,2}(k, q)&=\dfrac{\pi }{2 q^{2}}\,\delta''(k-q)-\dfrac{5  \pi}{2 q^{3}}\,\delta'(k-q)+\dfrac{3 \pi }{2 q^{4}}\,\delta(k-q), \\
\mathcal{I}^2_{4,3}(k, q)&=\dfrac{15\left(3 k^{4}+6 q^{2} k^{2}+7 q^{4}\right)}{16 k^{5} q^{4}}\ln \dfrac{k+q}{|k-q|}
\notag\\&
-\dfrac{45 k^{8}-30 k^{6} q^{2}-36 k^{4} q^{4}+190 k^{2} q^{6}-105 q^{8}}{8 q^{3} k^{4}(k-q)^{3}(k+q)^{3}}, \\
\mathcal{I}^2_{4,4}(k, q)&=-\dfrac{\pi }{2 q^{2}}\,\delta''(k-q)-\dfrac{ \pi}{q^{3}}\,\delta'(k-q)+\dfrac{9 \pi }{q^{4}}\,\delta(k-q).
\end{align}

%\cite{Gradshteyn:2007,Szalay:1997cc,Castorina:2017inr}
\newpage

\bibliographystyle{JHEP}
\bibliography{reference_library}

\end{document}